\documentclass[10pt,journal, twocolumn]{IEEEtran}
\usepackage{graphicx,times, amsmath, amsfonts,comment}
\usepackage{amssymb,epstopdf,bm,bbm,amsthm}
\usepackage[noend]{algorithmic}

\usepackage{multirow, mathtools}

\usepackage{epstopdf, soul, color}
\usepackage{algorithm, array}

\usepackage{relsize}
\newcommand{\beq}{\begin{equation}}
\newcommand{\eeq}{\end{equation}}

\newcommand*{\Scale}[2][4]{\scalebox{#1}{$#2$}}%
\theoremstyle{plain}

\begin{document}

%===========================================================================
\title{Non-Orthogonal Multiple Access (NOMA) for  Downlink Multiuser MIMO Systems: User Clustering, Beamforming, and Power Allocation}
%============================================================================
\author{\IEEEauthorblockN{Md Shipon Ali, Ekram Hossain, and Dong In Kim}\thanks{M. S. Ali and E. Hossain are with the Department of Electrical and Computer Engineering, at the University of Manitoba, Canada (Emails: alims@myumanitoba.ca, ekram.hossain@umanitoba.ca). D. I. Kim is with the School of Information and Communication
Engineering at the Sungkyunkwan University (SKKU), Korea (email: dikim@skku.ac.kr). The work was supported by a Discovery Grant form the Natural Sciences and Engineering Research Council of Canada (NSERC) and in part by the National Research Foundation of Korea (NRF) grant funded by the Korean government (MSIP) (2013R1A2A2A01067195) and (2014R1A5A1011478).} }
%\IEEEauthorblockA{Department of Electrical and Computer Engineering, University of Manitoba, Canada \\
%Email: \{ekram.hossain\}@umanitoba.ca}}
\maketitle

%========================================================================================
\begin{abstract}
We investigate the application of non-orthogonal multiple access (NOMA) with successive interference cancellation (SIC) in downlink multiuser multiple-input multiple-output (MIMO) cellular systems,  where the total number of receive antennas at user equipment (UE) ends in a cell is more than the number of transmit antennas at the base station (BS). We first dynamically group the UE receive antennas into a number of clusters equal to or more than the number of BS transmit antennas. A single beamforming vector is then shared by all the receive antennas in a cluster. We propose a linear 
beamforming technique in which all the receive antennas can significantly cancel the inter-cluster interference. On the other hand, the receive antennas in each cluster are scheduled on power domain NOMA basis with SIC at the receiver ends. For inter-cluster and intra-cluster power allocation, we provide dynamic power allocation solutions with an objective to maximizing the overall cell capacity. An extensive performance evaluation is carried out  for the proposed MIMO-NOMA system and the results are compared with those for conventional orthogonal multiple access (OMA)-based MIMO systems and other existing MIMO-NOMA solutions. The numerical results quantify the capacity gain of the proposed MIMO-NOMA model over MIMO-OMA and other existing MIMO-NOMA solutions.
\end{abstract}

%=======================================================================================

%=======================================================================================
\section{Introduction}
Non-orthogonal multiple access (NOMA) is being considered as a promising multiple access technology for fifth generation (5G) and beyond 5G (B5G) cellular systems that can significantly enhance system overall sprctral efficiency \cite{docomo2012}-\cite{msali2016}. The fundamental idea of NOMA  is to simultaneously serve multiple users over the same spectrum resources at the expense of inter-user interference. In contrast to the conventional orthogonal multiple access (OMA), where every user in a cell is served on exclusively allocated communication resources (time and frequency), NOMA superposes multiple users' message signals in power domain by exploiting their respective channel gain differences. Successive interference cancellation (SIC) is applied at the receiver ends for inter-user interference cancellation. 

On the other hand, multiple-input multiple-output (MIMO) communications with multiuser beamforming has been studied widely as a potential technology for achieving significant gains in the overall system throughput \cite{quentin2004}. In downlink multiuser MIMO, each UE is served by one or multiple beams depending on the number of BS transmit antennas and total number of receive antennas at the UEs  in the cell. The inter-beam interference can be completely eliminated when the number of transmit antennas is more than or equal to the number of receive antennas. In such a conventional multiuser MIMO system, each UE receiver is supported by individual beamforming vector which is orthogonal to the other receivers' channel gains. 

In a downlink multiuser MIMO-NOMA system, multiple receive antennas of different UEs with distinct channel gains are grouped into MIMO-NOMA clusters. All the users/receive-antennas in each cluster are scheduled on NOMA basis. The number of clusters is generally equal to the number of BS transmit antennas, i.e., the number of transmit beams. A single beam is utilized by all the receivers of a cluster which adopt NOMA. In case of more clusters than the BS transmit antennas, multiple clusters may share the same beam but utilize orthogonal spectrum resources, while each cluster individually is scheduled on a NOMA basis. However, since the number of UE receive antennas in downlink MIMO-NOMA system is much higher than the number of BS transmit antennas,  it results in strong inter-cluster interference. Therefore, multi-cluster beamforming is a key to reducing inter-cluster interference in MIMO-NOMA systems. In this paper, we propose a new multi-cluster zero-forcing beamforming (ZF-BF) technique for downlink multiuser MIMO-NOMA systems.

The MIMO-NOMA system in which beamforming is performed by using the channel information of the highest channel gain users of each cluster is defined as \textit{conventional MIMO-NOMA} in this paper, while orthogonal spectrum resource allocation among the users of each MIMO-NOMA cluster is defined as \textit{conventional MIMO-OMA or MIMO-OMA}. Also, a NOMA system with single antennas BS is referred as \textit{conventional NOMA}. In addition, we define the highest channel gain user of each MIMO-NOMA cluster as the \textit{cluster-head}, and each receive antenna in a cluster is defined as a \textit{user}.

%===========================================================================================
\textit{Notation}: In the reminder of this paper, bold face lower case and upper case characters denote column vectors and matrices, respectively. The operators $(\bullet)^H$, $(\bullet)^\dagger$, $|\bullet|$ and $||\bullet||_F$, correspond to the Hermitian transpose, the right pseudo-inverse, the absolute value and the Frobenius norm operations.

%========================================================================================
\subsection{Existing Research on Downlink MIMO-NOMA}
Recently,  application of NOMA in multiuser MIMO systems has drawn a significant attention of the researchers. The fundamentals of the combination of NOMA and MIMO in downlink transmission was studied in \cite{saito2013}, where a random beamforming was considered for  a $2$-user MIMO-NOMA cluster of single antenna UEs, and $2$ transmit antennas was assumed at the BS end. The authors in \cite{saito2013} utilized fixed power allocation strategies at the transmitter end, while a two-step interference cancellation method was considered at the UE receiver ends. In the first step, they used interference rejection combining (IRC) technique to suppress inter-beam/inter-cluster interference, while in the second step they used SIC for intra-beam/intra-cluster interference cancellation. 

In downlink transmission, the working principle of MIMO-NOMA and conventional NOMA are nearly similar. Each cluster in conventional NOMA is served by orthogonal spectrum resources, while the MIMO-NOMA clusters, the number of which is equal to the number of BS transmit antennas, use the same spectrum resources by utilizing MIMO principle. The power allocation within each cluster of MIMO-NOMA and conventional NOMA are exactly the same, and the dynamic power allocation is the key performance enabler in both cases. Regarding the dynamic power allocation in a NOMA cluster, the authors in \cite{msali2016} first formulated a convex optimization problem for intra-cluster power allocation, and then provided an optimal power allocation solution in closed-form for any cluster size.

In \cite{kim2013}, multiuser ZF-BF for downlink MIMO-NOMA was studied where each cluster contains only two single-antenna UEs (i.e., a $2$-user MIMO-NOMA system). The authors in \cite{kim2013} utilized conventional MIMO-NOMA model in which  ZF-BF was done by taking channel information of cluster-heads. The users are grouped into a cluster based on their channel gain correlations and differences with the channel gain of cluster-head. The authors also provided a user clustering algorithm which considers both of the channel correlation and gain differences among the users. The MIMO-NOMA system in \cite{kim2013} was only evaluated at highly correlated user cluster scenarios, but in practice, the radio channels for different UEs can be uncorrelated.

Another important study on downlink MIMO-NOMA  can be found in \cite{higuchi2015}, where random beamforming was performed at the BS and each beam is utilized by all the users of a cluster. The authors in \cite{higuchi2015} utilized spatial filtering to suppress the inter-cluster/inter-beam interference. The system level simulations in \cite{higuchi2015} were performed by considering only $2$ transmit antennas at the BS end, while up-to $4$-user MIMO-NOMA cluster with fixed power allocation strategies were evaluated. In addition, similar power allocation was assumed for each beam in \cite{higuchi2015}. The authors in \cite{higuch2013} utilzed the same system and signal model of \cite{higuchi2015} and proposed coordinated frequency block-dependent inter-beam power allocation to have distinct power levels for different beams. Moreover, random beamforming-based downlink MIMO-NOMA was also studied in \cite{lan2014} where the UE receivers use the minimum mean squired error (MMSE) technique to eliminate inter-beam interference. 

A study on downlink MIMO-NOMA was also presented in \cite{ding2016}, where the outage probability of the users of a MIMO-NOMA cluster was evaluated. The number of receive antennas at each UE end was considered to be equal to or more than the number of transmit antennas at the BS end. The authors simulated their model by considering up to $3$ antennas at the UE ends and the BS end, while up to $3$ UEs with fixed power allocation strategies were considered in each MIMO-NOMA cluster. In \cite{ding2016}, new precoding and decoding matrices were also proposed and it was claimed to obtain the ZF-BF characteristics. Another power allocation solution for a $2$-user downlink MIMO-NOMA cluster was studied in \cite{qi2015} where equal number of antennas are assumed at BS and each UE. A non-convex power allocation problem for $2$-user MIMO-NOMA was formulated in \cite{qi2015} and both optimal and sub-optimal solutions were proposed.

%=============================================================================================
\subsection{Motivation and Contributions}
NOMA is a cost-effective technology which can enhance cell spectral efficiency without requiring any additional infrastructure/resources. Although inter-user interference is the main obstacle for NOMA, efficient user clustering and power allocation can mitigate this and provide high spectral efficiency performance \cite{msali2016}. On the other hand, by using multiple antennas at the transmitter and receiver ends, MIMO technique can potentially multiply the spectral efficiency gain in proportion to the spatial multiplexing order. The inter-user interference in MIMO can be completely eliminated when the number of total receive antennas is equal to or less than the total transmit antennas in a cell. However, in MIMO-NOMA, the number of receive antennas is more than the number of transmit antennas, thus the resultant interference for each user is very high. Therefore, our motivation is to develop a ``robust" multiuser MIMO-NOMA system for downlink transmission which can minimize the net interference, and thus maximize the system capacity (or throughput). 

The major contributions of this paper can be summarized as follows: 

\begin{itemize}
\item We provide a novel system and signal model for downlink multiuser MIMO-NOMA. Our model is applicable to any MIMO-NOMA scenario in downlink transmission. 

\item We propose a new multi-cluster ZF-BF technique in which the precoding is performed by considering the equivalent channel gain of each MIMO-NOMA cluster, instead of that of any particular user. Each user in a MIMO-NOMA cluster cancels the inter-cluster interference by estimating its own {\em cluster's equivalent channel gain}. To estimate the {\em cluster's equivalent channel gain}, we introduce a decoding scaling weight factor for each particular user.

\item A low-complexity user clustering scheme is also proposed. The user clustering is performed with an objective to maximizing the  sum-spectral efficiency (or sum-throughput) in a cell by exploiting the NOMA and MIMO principles.   

\item We also provide a dynamic power allocation solution for downlink multiuser MIMO-NOMA. The power allocation is done in two steps: inter-cluster power allocation and intra-cluster power allocation. In inter-cluster power allocation, each transmit antenna gets power in proportion  to the number of users they serve, while the optimal power allocation solution developed in \cite{msali2016} is utilized for intra-cluster power allocation. 

\item Finally, we provide a comprehensive  performance evaluation of our proposed MIMO-NOMA model. The simulations are done for a wide rage of transmit antennas at the BS end and for different cluster sizes. Both of the correlated and uncorrelated Rayleigh fading channels are considered in the simulations.

\end{itemize}

%=========================================================================================
\subsection{Paper Organization}
The rest of the paper is organized as follows: Section~II presents the system model, assumptions,  and the signal model of our proposed downlink multiuser MIMO-NOMA system. Section~III discusses the proposed inter-cluster ZF-BF precoding and decoding techniques. Section IV and Section V, respectively, discuss the proposed user clustering scheme and dynamic power allocation solutions for downlink multiuser MIMO-NOMA system. Section VI evaluates the performance of the proposed solutions numerically, and Section~VII concludes the paper.

%========================================================
\section{System and Signal Models}
\subsection{System Model}
For a downlink multiuser MIMO-NOMA system, we consider a single cell BS equipped with $N_t$ transmit antennas for beamforming. The total number of UEs in the cell is $X$, where each UE can be equipped with one or more receive antenna(s). The total number of UE receive antennas in the cell is $L$, where $L > N_t$. The receive antennas are grouped into $N$ clusters, where $N \geq N_t$. In a cluster, no more than one user/receive-antenna belongs to one UE, while one user/receive-antenna only belongs to one cluster. However, if $N > N_t$, then multiple clusters may utilize the same BF vector but use orthogonal spectrum resources to each other, while users in each cluster are scheduled on a NOMA basis. For $N = N_t$, all the clusters use the same spectrum resources (full bandwidth) and each BF vector serves individual cluster. Each MIMO-NOMA cluster (\textit{we define each cluster as a MIMO-NOMA cluster}) consists of $K$ users/receive-antennas such that $\sum_{n=1}^N |K| = L$.

For the sake of simplicity, we consider that the number of clusters is equal to the number of transmit antennas, i.e., $N = N_t$.

%========================================================================================
\subsection{Signal Model}
Let us consider $\textbf{x} = [x_1\,\,\, x_2 \,\,\, x_3 \,\, \cdots \,\, x_N]^T\in\mathbb{C}^{N\times 1}$ to be the transmitted data  vector, where $x_n=\sum_{k=1}^{K}p_{n,k}s_{n,k}$ is the data stream for $n$-cluster in which $p_{n,k}$ and $s_{n,k}$ are the transmit power and message signal, respectively, for the $k$-th user in the $n$-th cluster. Let us also assume that the data vector is modulated by a beamforming precoding matrix $\textbf{M}\in\mathbb{C}^{N \times N}$, and then transmitted over the radio channel $\textbf{H} = [\textbf{H}_1^T \,\,\, \textbf{H}_2^T \,\,\, \textbf{H}_3^T \,\, \cdots \,\, \textbf{H}_N^T]^T \in\mathbb{C}^{L\times N}$, where $\textbf{H}_n \in\mathbb{C}^{K \times N}$ corresponds to the radio channel of all $K$ users of $n$-th cluster. Therefore, the transmitted superposed signal $\tilde{\textbf{x}}\in\mathbb{C}^{N\times 1}$ can be expressed as 
\begin{align}
\tilde{\textbf{x}} = \textbf{M}\textbf{x}.
\label{d1}
\end{align}
Now, we define $d_{n,k}\in\mathbb{C}$ to be the decoding scaling weight factor in which the received signal is multiplied prior to decode at $k$-th user of $n$-th cluster end. Thus, the received signal for $k$-th user of $n$-th cluster can be expressed as
\begin{align}
y_{n,k} = d_{n,k} [\textbf{h$_{n,k}$}\textbf{M}\textbf{x} + z_{n,k}] 
\label{d2}
\end{align}
where $\textbf{h}_{n,k} \in\mathbb{C}^{1 \times N}$ is the radio channel gain column vector for $k$-th user of $n$-th cluster, and $z_{n,k}\in\mathbb{C}$ represents circularly symmetric complex Gaussian noise with variance $\sigma^2$. However, if $\textbf{m}_n$ denotes the $n$-th column of the BF precoding matrix $\textbf{M}$, then \eqref{d2} can be expressed as follows:
\begin{align}
y_{n,k} &= d_{n,k}\textbf{h$_{n,k}$}\textbf{m$_n$}x_n + d_{n,k}\textbf{h$_{n,k}$}\sum\limits_{i=1, i\neq n}^N\textbf{m$_i$}x_i + d_{n,k}z_{n,k} \nonumber \\
&= d_{n,k}\textbf{h$_{n,k}$}\textbf{m}_n p_{n,k}s_{n,k} + d_{n,k}\textbf{h$_{n,k}$}\textbf{m}_n\sum\limits_{j=1, j\neq k}^{K} p_{n,j} s_{n,j} \nonumber \\
& \quad + d_{n,k}\textbf{h}_{n,k}\sum\limits_{i=1, i\neq n}^N\textbf{m}_i x_i + d_{n,k}z_{n,k}.
\label{d3}
\end{align}
In this downlink MIMO-NOMA system, the dynamic power allocation within each MIMO-NOMA cluster is performed in such a way that the higher channel gain user can perfectly decode and then suppress the intra-cluster interference from lower channel gain users. Thus, \eqref{d3} can be rewritten as follows:
\begin{align}
y_{n,k} = &d_{n,k}\textbf{h$_{n,k}$}\textbf{m}_n p_{n,k}s_{n,k} + d_{n,k}\textbf{h$_{n,k}$}\textbf{m}_n\sum\limits_{j=1}^{k-1} p_{n,j} s_{n,j} \nonumber \\
& + d_{n,k}\textbf{h}_{n,k}\sum\limits_{i=1, i\neq n}^N\textbf{m}_i x_i + d_{n,k}z_{n,k}.
\label{d4}
\end{align}
Therefore, the received signal-to-intra-cell interference-plus-noise ratio (SINR) for $k$-th user of $n$-th cluster can be expressed as follows:
\begin{align}
&{\rm SINR}_{n,k} = \nonumber \\
&\frac{|(d_{n,k}\textbf{h$_{n,k}$})\textbf{m}_n|^2 p_{n,k}}{ \Scale[.95]{ \underbrace{|(d_{n,k}\textbf{h$_{n,k}$})\textbf{m}_n|^2\sum\limits_{j=1}^{k-1} p_{n,j}}_\text{Intra-beam interference} + \underbrace{ \sum\limits_{i=1, i\neq n}^N |(d_{n,k}\textbf{h}_{n,k})\textbf{m}_i|^2 p_i}_\text{Inter-beam interference} + \underbrace{d_{n,k}z_{n,k}}_\text{Noise}} }
\label{d5}
\end{align}
where $p_i$ is the total transmit power for $i$-th cluster, and we assume that E$[|s_{i,j}|^2] = 1 \,\, \forall\, i,j$. The achievable throughput for $k$-th user of $n$-th cluster can be expressed as 
\begin{align}
\bar{R}_{n,k} = B \log_2\Bigg(1+\frac{g_{n,k}p_{n,k}}{g_{n,k}\sum\limits_{j=1}^{k-1} p_{n,j} +1}\Bigg)
\label{d6}
\end{align}
where $B$ is the total system bandwidth utilised by each transmit beam, and the resultant normalized channel gain $g_{n,k}$ could be defined as follows:
\begin{align}
g_{n,k}=\frac{|(d_{n,k}\textbf{h}_{n,k})\textbf{m}_n|^2}{\sum\limits_{i=1, i\neq n}^N |(d_{n,k}\textbf{h}_{n,k})\textbf{m}_i|^2 p_i + d_{n,k}z_{n,k}B}.
\label{d7}
\end{align} 
For this proposed downlink MIMO-NOMA system, the overall achievable cell system throughput can be expressed as
\begin{align} 
\bar{R}_{cell} = R\mathlarger{\mathlarger{\sum}}_{n = 1}^{N}\mathlarger{\mathlarger{\sum}}_{k = 1}^{K}B \log_2\Bigg(1+\frac{g_{n,k}p_{n,k}}{g_{n,k}\sum\limits_{j=1}^{k-1} p_{n,j} +1}\Bigg)
\label{d8}
\end{align}
where $\mathcal{U}_{n,k} \cap \mathcal{U}_{n^\prime,k} = \emptyset, \, \forall n\neq n^\prime \text{ and } \forall k $, $\mathcal{U}_{n,k}$ represents the $k$-th user in $n$-th cluster. The maximization of \eqref{d8} depends on three key factors: beamforming technique, user clustering approach, and inter-cluster and intra-cluster power allocations. 

The designs for the precoding matrix and decoding scaling weight factor will be discussed in Section III. In Section IV, we will discuss the user clustering approaches. In terms of power allocation, traditional water-filling approach for inter-cluster power allocation maximize the inter-cluster sum throughput, while the inverse water-felling approach is utilized by NOMA for intra-cluster power allocation. Therefore, a global dynamic power allocation may not possible in MIMO-NOMA system, rather we will utilize a two-step power allocation method which will be discussed in Section V.

%============================================================================================
\section{Beamforming in Downlink MIMO-NOMA}
In ZF-BF method, the inter-cluster interference could be completely removed when the number of total receive antennas is less than or equal to the number of total transmit antennas in a cell, thus each cluster contains maximum one receive antenna. In case of equal transmit and receive antennas, the ZF-BF can simply be achieved by taking the inversion of the receivers' channel gains. On the other hand, singular value decomposition (SVD) of the receivers' channel gains is used to achieve the ZF-BF in case of equal or less receive antennas than the transmit antennas in a cell. However, in a MIMO-NOMA system, the number of receive antennas is much higher than the number of transmit antennas, thus the multiuser ZF-BF may not be achieved. In this proposed downlink MIMO-NOMA system, we utilize a  precoding technique that was suggested in \cite{quentin2004}, where the actual channel matrix $\textbf{H}\in\mathbb{C}^{K\times N}$ for all $n$-th MIMO-NOMA clusters consisting of $K$ users are manipulated to form an equivalent channel matrix $\bar{\textbf{H}}\in\mathbb{C}^{1\times N}$ which provides compatible dimension for ZF-BF. Unfortunately, the performance of this  precoding technique not much better than the conventional MIMO-NOMA precoding. To improve the MIMO-NOMA performance with this  precoding technique, we introduce a decoding scaling weight factor which potentially reduces the inter-cluster interference while increasing the desired signal strength. The details of the precoding and decoding techniques will be discussed in the following subsections.

%=============================================================================================
\subsection{Precoding Matrix}
Let us consider the $n$-th MIMO-NOMA cluster which consists of $K$ users, then the radio channel matrix $\textbf{H}_n \in\mathbb{C}^{K \times N}$ can be expressed as follows:
\begin{align}
\textbf{H}_n = [\textbf{h}_{n,1}^T \,\,\, \textbf{h}_{n,2}^T \,\,\, \textbf{h}_{n,3}^T \,\, \cdots \,\, \textbf{h}_{n,K}^T]^T
\label{d9}
\end{align}
where $\textbf{h}_{n,k} \in \mathbb{C}^{1\times N}$ is the radio channel vector for the $k$-th user of $n$-th cluster. Now taking the SVD of the channel matrix $\textbf{H}_n$ we obtain
\begin{align}
\textbf{H}_n = \textbf{U}_n  {\bf\Sigma}_n \textbf{V}_n^H.
\label{d10}
\end{align}
In our proposed MIMO-NOMA system, each beamforming vector is utilized by one MIMO-NOMA cluster. Thus, the equivalent radio channel matrix $\bar{\textbf{H}}_n \in\mathbb{C}^{1 \times N}$ for the $n$-th cluster is obtained as follows:  
\begin{align}
\bar{\textbf{H}}_n = \textbf{U}_n^{(1)H} \textbf{H}_n
\label{d11}
\end{align}
where $\textbf{U}_n^{(1)H}\in \mathbb{C}^{1\times K}$ is the Hermitian transpose of the first column of $\textbf{U}_n$ of \eqref{d10}. Thus, the equivalent radio channel matrix $\bar{\textbf{H}} \in\mathbb{C}^{N \times N}$ consisting of all MIMO-NOMA clusters can be expressed as follows:
\begin{align}
\bar{\textbf{H}} = [\bar{\textbf{H}}_1^T \,\,\, \bar{\textbf{H}}_2^T \,\,\, \bar{\textbf{H}}_3^T \,\, \cdots \,\, \bar{\textbf{H}}_N^T]^T.
\label{d12}
\end{align}
Finally, the BF precoding matrix $\textbf{M} \in\mathbb{C}^{N \times N}$ can be obtained by taking the right pseudo-inverse of the equivalent matrix $\bar{\textbf{H}}$ as follows:
\begin{align}
\textbf{M} = \frac{\bar{\textbf{H}}^{\dagger}}{||\bar{\textbf{H}}^{\dagger}||_F} =  \Bigg[\frac{\bar{\textbf{m}}_1}{||\bar{\textbf{m}}_1||_F} \frac{\bar{\textbf{m}}_2}{||\bar{\textbf{m}}_2||_F} \cdots \frac{\bar{\textbf{m}}_N}{||\bar{\textbf{m}}_N||_F}\Bigg]
\label{d13}
\end{align}
where $||\bullet||_F$ represents the Frobenius norm, and $\bar{\textbf{m}}_n$ indicates the $n$-the column of the pseudo-inverse of $\bar{\textbf{H}}$.

%================================================================================================
\subsection{Decoding Scaling Weight Factor}
In this proposed downlink MIMO-NOMA system, the users/receive-antennas are clustered in such a way that each cluster contains maximum distinctive channel gain users (the details of user clustering will be presented in Section IV). Also, within each cluster, the users are sorted according to the ascending order of their channel gains, thus the strongest user is the $1$st user (\textit{which is also defined as cluster-head}) and the weakest user is the $K$-th user of a MIMO-NOMA cluster consisting of $K$ users. On the other hand, according to the properties of the SVD of a matrix $\textbf{A}$, the elements of first column of the left singular matrix $(\textbf{U}^{(1)})$ is nearly proportional to the normalized value of the corresponding rows of the matrix $\textbf{A}$. If the channel gain differences between the cluster-head and other users of a MIMO-NOMA cluster are sufficiently large, the equivalent channel gain is then nearly similar to the cluster-head's channel gain. Therefore, in such a channel condition, the cluster-head of each MIMO-NOMA cluster can almost completely cancel inter-cluster interference. All the cluster members except the cluster-head can achieve a better inter-cluster interference cancellation if they estimate their equivalent cluster channel or cluster-head's channel at their ends. To obtain the best estimation of cluster-head's channel for $k$-th user of $n$-th MIMO-NOMA cluster, his own channel need to be scaled up by the following weight factor prior to the decoding:
\begin{align}
d_{n,k} = \frac{\textbf{U}_{n,1}^{(1)}}{\textbf{U}_{n,k}^{(1)}},  \,\, \quad \forall n \text{ and } \forall k
\end{align}
where $\textbf{U}_{n,k}^{(1)}$ is the $k$-th row of first column of the left singular matrix ${\textbf{U}_n}$ of $n$-th MIMO-NOMA cluster's channel matrix $\textbf{H}_n$. In this proposed downlink MIMO-NOMA system, the BS sends the decoding scaling weight factor to the specific users before it starts  sending the data streams. This is easy for MIMO-NOMA system to send such information before sending data streams, since the NOMA with dynamic power allocation requires to share the allocated transmit power information to the participating NOMA users in order to perform SIC  at the receiver ends.

%============================================================================================
\section{Dynamic User Clustering in Downlink MIMO-NOMA}
As mentioned in \cite{msali2016}, the optimal user clustering for conventional NOMA system requires an exhaustive search among all the users in a cell. That is, for every single user, it is required to consider all possible combinations of user grouping. In downlink transmission, the working principle of MIMO-NOMA and conventional NOMA are nearly similar. Each cluster of a conventional NOMA is served by orthogonal spectrum resources, while MIMO-NOMA clusters equal to the number of BS antennas use same spectrum resources by utilizing MIMO principle. Therefore, the computational complexity of optimal user clustering for downlink MIMO-NOMA is also extremely high, and thus may not be affordable for practical systems with a moderate to large number of transmit and receive antennas.

In this section, we propose a low-complexity sub-optimal user clustering scheme for downlink MIMO-NOMA system. The proposed scheme exploits the channel gain differences and correlations among the NOMA users, and targets to enhance the overall sum-spectral efficiency in the cell. Note that, in this paper, the channel gain correlation between two users refers to their Rayleigh fading gain correlation. By utilizing the channel gain differences, a low-complexity user clustering scheme was also proposed in \cite{msali2016} for a conventional NOMA system. However, the user clustering of MIMO-NOMA is little bit different from that of conventional NOMA. Along with the channel gain differences and correlations among the NOMA users, the precoding and decoding techniques of MIMO system play a vital role in MIMO-NOMA user clustering. 

%=============================================================================================
\subsection{Key Issues in User Clustering for Downlink MIMO-NOMA}
With an objective of sum-throughput maximization in a cell, the key factors that need to be considered for user clustering in downlink MIMO-NOMA system are as follows.

\renewcommand{\labelitemi}{$\blacksquare$}
\begin{itemize}
\item A mentioned in \cite{msali2016}, the cluster-head (the highest channel gain user) of a downlink NOMA cluster can completely cancel intra-cluster interference, and thus achieves maximum throughput gain. Therefore, a key point to maximizing the overall system capacity (or spectral efficiency) is to ensure that the high channel gain users in a cell are selected as the cluster-heads of different MIMO-NOMA clusters. 

\item As discussed in Section III, the equivalent channel gain of this proposed MIMO-NOMA cluster is almost similar to the channel gain of the cluster-head when the  differences in channel gains of the cluster-head and the other users  are sufficiently high. In addition, the proposed decoding scaling weight factor can sufficiently improve the desired signal and eliminate the inter-cluster interference for the remaining users of a MIMO-NOMA cluster when their channel gains are significantly much lower than that of the cluster-head. Therefore, the second key point of downlink MIMO-NOMA user clustering is to make the cluster-head much distinctive from the remaining users of the cluster. This point also indicates that all the users under a MIMO cell may not be able to use NOMA. 

\item In both the proposed and conventional MIMO-NOMA clusters, the cluster members with more correlated  channel gains with the cluster-head can effectively eliminate the inter-cluster interference, and thus improve the resultant capacity gain. In general, the channel gains between two users would be correlated when they are closely located, i.e., difference in their channel gains is small. However, the  beamforming technique proposed for the MIMO-NOMA system requires high channel gain differences between the cluster-head and the other users of a MIMO-NOMA cluster. Therefore, both the channel gain differences and correlations need to be considered while  forming  downlink MIMO-NOMA clusters.

%\item In both the proposed and conventional MIMO-NOMA clusters, the cluster members with more correlated channel gains with the cluster-head (i.e., when the channel gain differences are small) can effectively eliminate the inter-cluster interference, and thus improve the resultant capacity gain.  Therefore, both the channel gain differences and correlations need to be considered while  forming  downlink MIMO-NOMA clusters. 
\end{itemize}

%=================================================================================================================
\subsection{User Clustering Algorithm}
Let us consider a downlink MIMO-NOMA system where the number of transmit antennas at the BS is $N_t$. The total number of receive antennas at all UEs in the cell is $L$. Here, we first provide a low-complexity MIMO-NOMA user clustering algorithm in which the number of clusters $N = N_t$ and each cluster contains $K$ (equal) number of receive antennas of different UEs. Afterwards, a more sophisticated version of MIMO-NOMA user clustering algorithm will be presented. In addition, we define $\rho$ as the minimum correlation coefficient for which two users' Rayleigh fading channel gains are considered to be correlated. Moreover, we use $(\mathcal{U}_i \neq \mathcal{U}_j)$ to indicate that both of the $i$-th and $j$-th receive antennas in the cell do not belong to the same UE. The low-complexity MIMO-NOMA user clustering algorithm is given in \textbf{Algorithm 1}.
\rule[.1ex]{\linewidth}{1.5pt}
{\textbf{Algorithm $1$: User Clustering for Downlink MIMO-NOMA}
\rule[.8ex]{\linewidth}{.5pt}
\begin{enumerate}
\item[\textbf{1.}] \textbf{Sort users according to the ascending channel gain:} \\
$h_1 \geq h_2 \geq  \cdots \geq h_L$, $h_i = i$-th user's channel gain. 
\vspace{6pt}\item[\textbf{2.}] \textbf{Select number of clusters and cluster-heads:} \\
Number of MIMO-NOMA clusters is $N$, and the $n$-th higher channel gain user in a cell is the cluster-head of $n$-th MIMO-NOMA cluster.
\vspace{6pt}
\hspace{6pt}\item[\textbf{3.}] \textbf{Include second users into each cluster:} \\
Initiate user sets: $\mathcal{A}= \{1,2,\cdots,N\}$, $\mathcal{B}= \{N+1,N+2,\cdots,2N\}, \mathcal{R}_{i,j} = \text{correlation coefficient }$ between $h_i$ and $h_j$.
\vspace{5pt}
\item[(\textbf{\textit{a}})]\textbf{\textit{Clustering of users with correlated channel gains:}}\\
for $i=1:N$ \\
\text{\hspace{5 pt}}for $j=N+1:2N$\\
\text{\hspace{10 pt}}if $\big(\mathcal{R}_{i,j} > \mathcal{R}_{i,k} \geq \rho, \forall k\neq j \in \mathcal{B}\big) \text{ AND }\big(\mathcal{U}_i \neq \mathcal{U}_j\big)$ \\
\text{\hspace{20 pt}}include $j$-th user into $i$-th cluster, \\
\text{\hspace{20 pt}}update $\mathcal{A} \leftarrow \mathcal{A} - \{i\}$, $\mathcal{B} \leftarrow \mathcal{B} - \{j\}.$\\
\text{\hspace{10 pt}}end \\
\text{\hspace{5 pt}}end \\
end
\vspace{5pt}
\item[(\textbf{\textit{b}})]\textbf{\textit{Clustering of users with uncorrelated channel gains:}}\\
\text{for } $i=1:N$ \\
\text{\hspace{5 pt}}for $j=N+1:2N$ \\
\text{\hspace{10 pt}}if $\big(i \in \mathcal{A} \big)\, \text{AND}\, \big(j \in \mathcal{B}\big)\, \text{AND}\,\big(\mathcal{U}_i \neq \mathcal{U}_j\big)$\\
\text{\hspace{20 pt}}include $j$-th user into $i$-th cluster, \\
\text{\hspace{20 pt}}update $\mathcal{A} \leftarrow \mathcal{A} - \{i\}$, $\mathcal{B} \leftarrow \mathcal{B} - \{j\}$.\\
\text{\hspace{10 pt}}end \\
\text{\hspace{5 pt}}end \\
end \\

\item[\textbf{4.}] \textbf{Include $k$-th users into each cluster:} \\
Repeat Step $3$, \\
while $3\leq k \leq K$\\
\text{\hspace{15 pt}}set $i = 1:N$, $j = kN-N+1:kN$, $\mathcal{A}= \{1, 2,\cdots,\\ \text{\hspace{15 pt}} N\}$, $\mathcal{B}= \{kN-N+1,kN-N+2,\cdots,kN\}$. \\
end
\end{enumerate}
\rule[1ex]{\linewidth}{1.5pt}

In \textbf{Algorithm 1}, it is assumed that all the users in a downlink MIMO cellular system can utilize NOMA-based resource allocation. However, the lower channel gain users of a NOMA cluster usually experience high intra-cluster interference.  The lower channel gain users in a MIMO-NOMA cluster also experience strong inter-cluster interference which results in a low SINR. Although our proposed MIMO-NOMA system can significantly cancel inter-cluster interference for lower channel gain users, the performances of these users deteriorate as their channel gains become closer to that of the cluster-head. Therefore, only a portion of the users may use NOMA in a MIMO system. In addition, a higher order MIMO-NOMA system (i.e., a system with more users in a cluster) requires more transmit power for the lowest channel gain user of that cluster, while the transmit power budget per antenna in MIMO system is generally limited. Therefore, a higher order clustering may not to be possible in downlink MIMO-NOMA system. On the other hand, to maximize the cell sum-spectral efficiency, the users utilizing MIMO-NOMA  should be clustered in such a way that each cluster contains users with the maximum channel gain differences, in case of uncorrelated channel gains.

Based on the aforementioned practical considerations, a sophisticated user clustering algorithm for $2$-user clusters for the proposed MIMO-NOMA system is given in \textbf{Algorithm $2$}. Note that the clustering process for $3$-user and higher order MIMO-NOMA systems would be similar to that given by \textbf{Algorithm $2$}.
\rule[.1ex]{\linewidth}{1.5pt}
{\textbf{Algorithm $2$: User Clustering for $2$-User Downlink MIMO-NOMA.\\}
\rule[.8 ex]{\linewidth}{.5 pt}
\begin{enumerate}
\item[\textbf{1.}] \textbf{Input, Initialization and Assumption:} \\
\textit{Input:} Tx antennas $N_t$, Rx antennas $L$, channel gain $h_i$. \\
\textit{Initialize:} $\mathcal{C}_{i,j} = 1$, if $i$-th and $j$-th users can cluster, $\mathcal{C^\prime}(t) = [i,j]$, if $i$-th and $j$-th users form $t$-th cluster.\\
\textit{Assumption:} In uncorrelated channel gain, if $j$-th user can cluster with $i$-th user, it also can cluster with all $i^\prime$ users such that $h_{i^\prime} > h_i.$
\vspace{6pt}
\item[\textbf{2.}] \textbf{Sort users according to the ascending channel gain:} \\
$h_1 \geq h_2 \geq  \cdots \geq h_L$, $h_i = i$-th user's channel gain. 
\vspace{6pt}
\item[\textbf{3.}] \textbf{Initial selection of the number of clusters:} \\
If $N_t<L\leq 2N_t$, then initial clusters are $N = N_t$, elseif $L>2N_t$, then initial clusters are $N = \lceil{L/2}\rceil.$
\vspace{6pt}
\hspace{6pt}\item[\textbf{4.}] \textbf{User selection for clustering:} \\
Initialise user set: $\mathcal{A}= \{1,2,\cdots,N\}, \, \mathcal{B}= \{N+1, N+2,\cdots,L\}$. Also, initialize two arrays $\mathcal{A^\prime} \text{ and } \mathcal{B^\prime}$ which contain the users, who are able to use NOMA, from set $\mathcal{A} \text{ and } \mathcal{B}$, respectively. Define $\mathcal{R}_{i,j} =$ correlation coefficient between $h_i$ and $h_j$. $\vspace{5 pt} \text{ Initialise } t =  1, \, t^\prime = 1.$
\item[(\textbf{\textit{a}})]\textbf{\textit{Clustering of users with correlated channel gains:}}\\
$\text{for } i=1:N \\
\text{\hspace{5 pt}for } j=N+1:L \\
$\text{\hspace{10 pt}}if $ \Scale[.8]{\big(\mathcal{R}_{i,j} > \mathcal{R}_{i,k} \geq \rho, \forall k\neq j \in \mathcal{B}\big)\,\text{AND}\, \big(\mathcal{C}_{i,j}=1\big)\,\text{AND}\,\big(\mathcal{U}_i \neq \mathcal{U}_j\big)}$ \\
\text{\hspace{20 pt}}$\mathcal{C}^\prime(t) = [i,j], \, t \leftarrow t+1,$ \\
\text{\hspace{20 pt}}update $\mathcal{A} \leftarrow \mathcal{A} -\{ i\}, \, \mathcal{B} \leftarrow \mathcal{B} - \{j\}$.\\
\text{\hspace{10 pt}}end \\
\text{\hspace{5 pt}}end \\
\vspace{5 pt}end 
\item[(\textbf{\textit{b}})]\textbf{\textit{Clustering of users with uncorrelated channel gains:}}\\
set $i = N$, $j = L$ \\
while $j\geq N +1$ \\
\text{\hspace{5 pt}}while $i\geq 1$ \\
\text{\hspace{10 pt}}if $\Scale[.92]{\big(i \in \mathcal{A} \big)\, \text{AND}\, \big(j \in \mathcal{B}\big)\, \text{AND}\, \big(\mathcal{C}_{i,j} = 1\big)\,\text{AND}\,\big(\mathcal{U}_i \neq \mathcal{U}_j\big)}$ \\
\text{\hspace{20 pt}}$\mathcal{A^\prime}(t^\prime) = i, \, \mathcal{B^\prime}(t^\prime) = j,$ \\
\text{\hspace{20 pt}}update: $\mathcal{A} \leftarrow \mathcal{A} - \{ i\}, \, \mathcal{B} \leftarrow \mathcal{B} - \{ j\}, $\\
\vspace{5 pt}\text{\hspace{20 pt}}$t^\prime \leftarrow t^\prime+1, \, j \leftarrow j-1, \, i \leftarrow i-1.$ \\
\text{\hspace{0 pt}}elseif $\Scale[.90]{\big(i \in \mathcal{A} \big)\, \text{AND}\, \big(j \in \mathcal{B}\big)\, \text{AND}\, \big(\mathcal{C}_{i,j} \neq 1\big)\,\text{AND}\,\big(\mathcal{U}_i \neq \mathcal{U}_j\big)}$ \\
\vspace{5 pt}\text{\hspace{20 pt}}update: $\mathcal{A} \leftarrow \mathcal{A} - \{i\}, \, i \leftarrow i-1$. \\
\text{\hspace{0 pt}}elseif $\Scale[.90]{\big(j \not\in \mathcal{B} \big)\, \text{AND}\, \big(\mathcal{B} \neq \emptyset\big)\, \text{AND} \, \big(\mathcal{A} \neq \emptyset \big)\,\text{AND}\,\big(\mathcal{U}_i \neq \mathcal{U}_j\big)}$ \\
\text{\hspace{20 pt}}update: $\mathcal{A} \leftarrow \mathcal{A} - \{i\}, \, \mathcal{B} \leftarrow \mathcal{B} - \{j\},$ \\
\vspace{5 pt}\text{\hspace{20 pt}}$i \leftarrow i-1, \, j \leftarrow j-1$. \\
\text{\hspace{0 pt}}elseif $\Scale[.90]{\big(i \not\in \mathcal{A} \big)\, \text{AND}  \, \big(j \in \mathcal{B}\big)\, \text{AND}  \,\big(\mathcal{A} \neq \emptyset\big)\,\text{AND}\,\big(\mathcal{U}_i \neq \mathcal{U}_j\big)}$ \\
\vspace{5 pt}\text{\hspace{20 pt}}update: $\mathcal{A} \leftarrow \mathcal{A} - \{i\}, \, i \leftarrow i-1$.  \\
\text{\hspace{0 pt}}elseif $\Scale[.90]{\big(i \in \mathcal{A} \big)\, \text{AND}  \, \big(j \not \in \mathcal{B}\big)\, \text{AND}  \,\big(\mathcal{B} = \emptyset\big)\,\text{AND}\,\big(\mathcal{U}_i \neq \mathcal{U}_j\big)}$ \\
\text{\hspace{20 pt}}for $k = 1: t^\prime -2$\\
\text{\hspace{35 pt}}$\mathcal{A^\prime}(k) = \mathcal{A^\prime}(k+1), \, \mathcal{B^\prime}(k) = \mathcal{B^\prime}(k+1)$\\
\vspace{5 pt}\text{\hspace{20 pt}}end \\
\text{\hspace{20 pt}}$\mathcal{A^\prime}(k+1) = i, \, \mathcal{B^\prime}(k+1) = j,$\\
\text{\hspace{20 pt}}update: $\mathcal{A} \leftarrow \mathcal{A} - \{ i\}, \,  i \leftarrow i-1, \,  k \leftarrow k+1$. \\
\text{\hspace{10 pt}}end \\
\text{\hspace{5 pt}}end \\
end \\

\item[\textbf{5.}] \textbf{Final MIMO-NOMA clusters:} \\
while $k \geq 1$ \\
\text{\hspace{15 pt}}$\mathcal{C}^\prime(t) = [\mathcal{A^\prime}(k), \, \mathcal{B^\prime}(k)]$, \\
\text{\hspace{15 pt}}update: $t \leftarrow t+1, \,  k \leftarrow k-1.$ \\
end 
\end{enumerate}
\rule[1.5ex]{\linewidth}{1.5pt}

The Step $4$(a) in \textbf{Algorithm $2$} indicates that, for clustering, the users, which have higher channel gain correlations with the cluster-head, can have a more relaxed requirement of the maximum gain differences. However, in case of uncorrelated channel gain,  for every cluster of a downlink multiuser MIMO-NOMA system, the Step $4$(b) in \textbf{Algorithm $2$} ensures maximum channel gain differences among the cluster-head and the other users of a cluster. It is important to note that the BS first utilizes \textbf{Algorithm $2$} for higher order user clustering, and then successively utilizes that to make the lower order user clustering for the users who are unable to form higher order MIMO-NOMA clusters.

%========================================================================================
\section{Dynamic Power Allocation in Downlink MIMO-NOMA}

For the proposed MIMO-NOMA system, we utilize a two-step power allocation method. Since each beam is utilized by all the users of a cluster,  the total BS transmit power is divided into the number of transmit beams such that the transmit power for a beam is proportional to the number of users served by that beam. If all the transmit beams serve equal number of users (same cluster size), then the beam transmit powers are equally allocated. This approach can be said as nearly optimal since each MIMO-NOMA cluster contains users with similar channel gain distinctness. However, the users in each cluster are scheduled according to the NOMA principle, and thus the intra-cluster/intra-beam dynamic power allocation is crucial. For intra-cluster power allocation, we provide a dynamic power allocation solution under the constraints of the beam transmit power budget which is determined at inter-beam/inter-cluster power allocation step, minimum rate requirements of the users of the subjected MIMO-NOMA cluster, and the requirement of the minimum power differences among the NOMA received signals to perform SIC at the receiver ends. 

Let us consider one MIMO-NOMA cluster, say the $n$-th cluster, in which the resultant normalized channel gains  defined by \eqref{d7} of all the $K$ users are $g_{n,1}, g_{n,2},\cdots,g_{n,K}$, respectively, where $g_{n,1} > g_{n,2} > \cdots > g_{n,K}$. The respective guaranteed throughput requirements of all $K$ users are assumed as $R_{n,1}, R_{n,2},\cdots,R_{n,K}$, where $R_{n,k}>0,\,\forall n \text{ and } \forall k $. If the number of MIMO-NOMA clusters is equal to the number of BS transmit antennas, then each beam utilizes the total system bandwidth $B$ to serve the users of a cluster. If $p_{n,1}, p_{n,2},\cdots,p_{n,K}$ are the transmit powers for $1$st, $2$nd, and $K$-th users of the $n$-th MIMO-NOMA cluster, respectively, then the optimal intra-cluster power allocation problem for $n$-th MIMO-NOMA cluster can be formulated as follows:

\begin{align} 
&\quad\underset{p}{\text{maximize}} \mathlarger{\mathlarger{\sum}}_{k = 1}^{K}B \log_2\Bigg(1+\frac{g_{n,k}p_{n,k}}{g_{n,k}\sum\limits_{j=1}^{k-1} p_{n,j} +1}\Bigg)
\label{v.1}
\\ 
&\text{ s.t.:}\enspace
%\nonumber\\ 
%&\quad\quad 
\textbf{C$_1$:}\,\,\sum\limits_{k = 1}^{K}p_{n,k} \leq p_n \nonumber \\ 
&\qquad \, \textbf{C$_2$:}\,\,\, B \log_2\Big(1+\frac{g_{n,k}p_{n,k}}{g_{n,k}\sum\limits_{j=1}^{k-1} p_{n,j} +1}\Big) \geq R_{n,k}, \, \forall k \nonumber\\
&\qquad\, \textbf{C$_3$:}\,\,\, \Big(p_{n,k} - \sum\limits_{j = 1}^{k-1}p_{n,j}\Big)g_{n,k-1} \geq p_{tol}, \, \forall k \neq 1 \nonumber
\end{align}
where $p_{tol}$ indicates the minimum received power differences among different users' signals of a MIMO-NOMA cluster to perform SIC. The optimization problem in \eqref{v.1} is similar to the dynamic power allocation in downlink NOMA for single antenna BS presented in \cite{msali2016}. By using the closed-form solution of \cite{msali2016}, we can find the dynamic intra-cluster power allocation solution for our proposed downlink multiuser MIMO-NOMA system. Therefore, the optimal power allocation for $1$st user of $n$-th MIMO-NOMA cluster is obtained as follows: 
\begin{align*}
p_{n,1} = \frac{p_n}{\underset{j\not\in \mathcal B^\prime}{\prod\limits_{j=2}^{K}}\varphi_{n,j} \underset{j\in \mathcal B^\prime}{\prod\limits_{j=2}^{K}}2} - \underset{j\not\in \mathcal B^\prime}{\mathlarger{\mathlarger{\sum}}_{j=2}^{K}}\frac{(\varphi_{n,j} - 1)}{g_{n,j} \underset{j^\prime\not\in \mathcal B^\prime}{\prod\limits_{j^\prime=2}^{j}}\varphi_{n,j^\prime} \underset{j^\prime\in \mathcal B^\prime}{\prod\limits_{j^\prime=2}^{j}}2} - \\
\underset{j\not\in \mathcal C^\prime}{\mathlarger{\mathlarger{\sum}}_{j=2}^{K}}\frac{p_{tol}}{2g_{n,j-1} \underset{j^\prime\not\in \mathcal B^\prime}{\prod\limits_{j^\prime=2}^{j-1}}\varphi_{n,j^\prime} \underset{j^\prime\in \mathcal B^\prime}{\prod\limits_{j^\prime=2}^{j-1}}2}.
\end{align*}
On the other hand, the optimal power allocation solution for $k$-th user ($\forall k \neq 1$) of the $n$-th MIMO-NOMA cluster could be expressed as follows: \vspace{6 pt}\\
(i) If $k\not\in \mathcal B^\prime$, then
\begin{align*}
p_{n,k} = &\Bigg[\frac{p_n}{\underset{j\not\in \mathcal B^\prime}{\prod\limits_{j=k}^{K}}\varphi_{n,j} \underset{j\in \mathcal B^\prime}{\prod\limits_{j=k}^{K}2}} - \underset{j\not\in \mathcal B^\prime}{\mathlarger{\mathlarger{\sum}}_{j=k}^{K}}\frac{(\varphi_{n,j} - 1)}{g_{n,j} \underset{j^\prime\not\in \mathcal B^\prime}{\prod\limits_{j^\prime=k}^{j}}\varphi_{n,j^\prime} \underset{j^\prime\in B^\prime}{\prod\limits_{j^\prime=k}^{j}}2}-\\
&\underset{j\not\in \mathcal C^\prime}{\mathlarger{\mathlarger{\sum}}_{j=k}^{K}}\frac{p_{tol}}{2g_{n,j-1} \underset{j^\prime\not\in \mathcal B^\prime}{\prod\limits_{j^\prime=k}^{j-1}}\varphi_{n,j^\prime} \underset{j^\prime\in \mathcal B^\prime}{\prod\limits_{j^\prime=k}^{j-1}}2} + \frac{1}{g_{n,k}}\Bigg]\times (\varphi_{n,k} - 1).
\end{align*}
(ii) If $k\in \mathcal B^\prime$, then
\begin{align*}
p_{n,k} = &\frac{p_n}{\underset{j\not\in \mathcal B^\prime}{\prod\limits_{j=k}^{K}}\varphi_{n,j} \underset{j\in \mathcal B^\prime}{\prod\limits_{j=k}^{K}}2} - \underset{j\not\in \mathcal B^\prime}{\mathlarger{\mathlarger{\sum}}_{j=k}^{K}}\frac{(\varphi_{n,j} - 1)}{g_{n,j} \underset{j^\prime\not\in \mathcal B^\prime}{\prod\limits_{j^\prime=k}^{j}}\varphi_{n,j^\prime} \underset{j^\prime\in \mathcal B^\prime}{\prod\limits_{j^\prime=k}^{j}}2}- \\
&\underset{j\not\in \mathcal C^\prime}{\mathlarger{\mathlarger{\sum}}_{j=k}^{K}}\frac{p_{tol}}{2g_{n,j-1} \underset{j^\prime\not\in \mathcal B^\prime}{\prod\limits_{j^\prime=k}^{j-1}}\varphi_{n, j^\prime} \underset{j^\prime\in \mathcal B^\prime}{\prod\limits_{j^\prime=k}^{j-1}}2} + \frac{p_{tol}}{g_{n,k-1}}
\end{align*}
where $\mathcal{B}^\prime$ and $\mathcal{C}^\prime$ are the complementary set of the minimum rate requirements of the users of $n$-th MIMO-NOMA cluster and the SIC constraints, respectively, which were defined in \cite{msali2016}. In addition, $\varphi_{n,k} = 2^{\frac{R_{n,k}}{B}}$ was also defined in \cite{msali2016}.

%==================================================================================
\section{Numerical Results and Discussion}
\subsection{Simulation Assumptions}
In this section, we provide simulation results to demonstrate the throughput/spectral-efficiency gain of the proposed MIMO-NOMA system and also compare the results with those of conventional MIMO-OMA and conventional MIMO-NOMA (i.e., particular user specific beamforming MIMO-NOMA) systems. The major simulation parameters are listed in Table~\ref{MIMO-NOMA parameters}, which follow the simulation assumptions in 3GPP-LTE \cite{3GPP2006}. 

\begin{table} [H]
\centering
\caption{Simulation parameters}
\label{MIMO-NOMA parameters}
\begin{tabular}{|c|c|}
\hline
Parameter         & Value                          \\
\hline \hline
Inter-site distance      & $0.6$ Km                       \\
\hline
System effective bandwidth, $B$     & $8.64$ MHz                       \\
\hline
Bandwidth of one resource block & $180$ kHz                             \\
\hline 
Number of available resource units & $48$                              \\
\hline 
Number of antennas at BS end, $N_t$     & $3, 5, 10$                     \\
\hline
Number of antennas at each UE end    & $1$                     \\
\hline 
Number of users/Rx-antennas in each cluster     & $2, 3, 4$                     \\
\hline
BS per antenna transmit power budget, $p_n$     & $43$ dBm                       \\
\hline
Antenna gain at BS/UE end    & $0$ dBi                      \\
\hline
SIC receiver's detection threshold, $p_{tol}$    & $10$ dBm  \\
\hline
Path-loss exponent, $\alpha$   & $4$  \\
\hline
Receiver noise density, $N_0$    & $-169$ dBm/Hz  \\
\hline                     
\end{tabular}
\end{table}

The radio channel is assumed to be the product of free-space path loss and Rayleigh fading with zero mean and unit variance. In these single-cell simulations, a single sector BS consisting of the whole hexagon is considered, where the BS is located at a corner of the cell areas. The  cluster-heads are assumed to be randomly distributed around the BS. On the other hand, the random distributions of the other users of each MIMO-NOMA cluster are measured in terms of the \textit{cell-edge coverage distance}. As an example, for an inter-site distance of 600 meter, 150m cell-edge coverage distance means that the users in a cell (except the cluster-head) are distributed within 450m to 600m  from the BS. Perfect CSI is assumed to be available at BS end. All the simulations are done for a single transmission time interval (TTI), that is, we only consider the instantaneous channel gains for a particular TTI.  However, these instantaneous channel gains are averaged over twenty thousands of channel realizations. In addition, the number of MIMO-NOMA clusters is assumed to be equal to the number of BS transmit antennas, thus all the clusters use full spectrum resources by utilizing the MIMO principle. Fig.~\ref{fig:vi.a} demonstrates the simulation model for a $3$-user MIMO-NOMA setup. 

For MIMO-OMA, the spectral efficiency is calculated by considering orthogonal spectrum resource allocation among the users of each cluster and the transmit power allocation for each user of a cluster is  proportional to the spectrum resource allocated to it. Also, before transmitting the data streams, the BS provides the decoding scaling weight factor and the transmit power information to the participating users in a cluster. Moreover, the users are clustered according to the sophisticated user clustering procedure presented in \textbf{Algorithm $2$}. For a particular simulation, we consider same size of all MIMO-NOMA clusters, and the guaranteed throughput requirement of a particular user is same for all clusters in that simulation setup. 
As an example, $R_{1,2} = R_{2,2} = R_{3,2}$ is the throughput requirement for the second higher channel gain user of clusters 1, 2, and 3. For brevity, we  use $R_2$, instead of  $R_{1,2} = R_{2,2} = R_{3,2}$. Similarly, $R_3$ indicates the guaranteed  throughput requirement for the third higher channel gain user of all cluster in a particular simulation.
The throughput/spectral-efficiency calculation is performed by using Shanon's capacity formula. 

\begin{figure}[h]
\begin{center}
	\includegraphics[width=3.5 in]{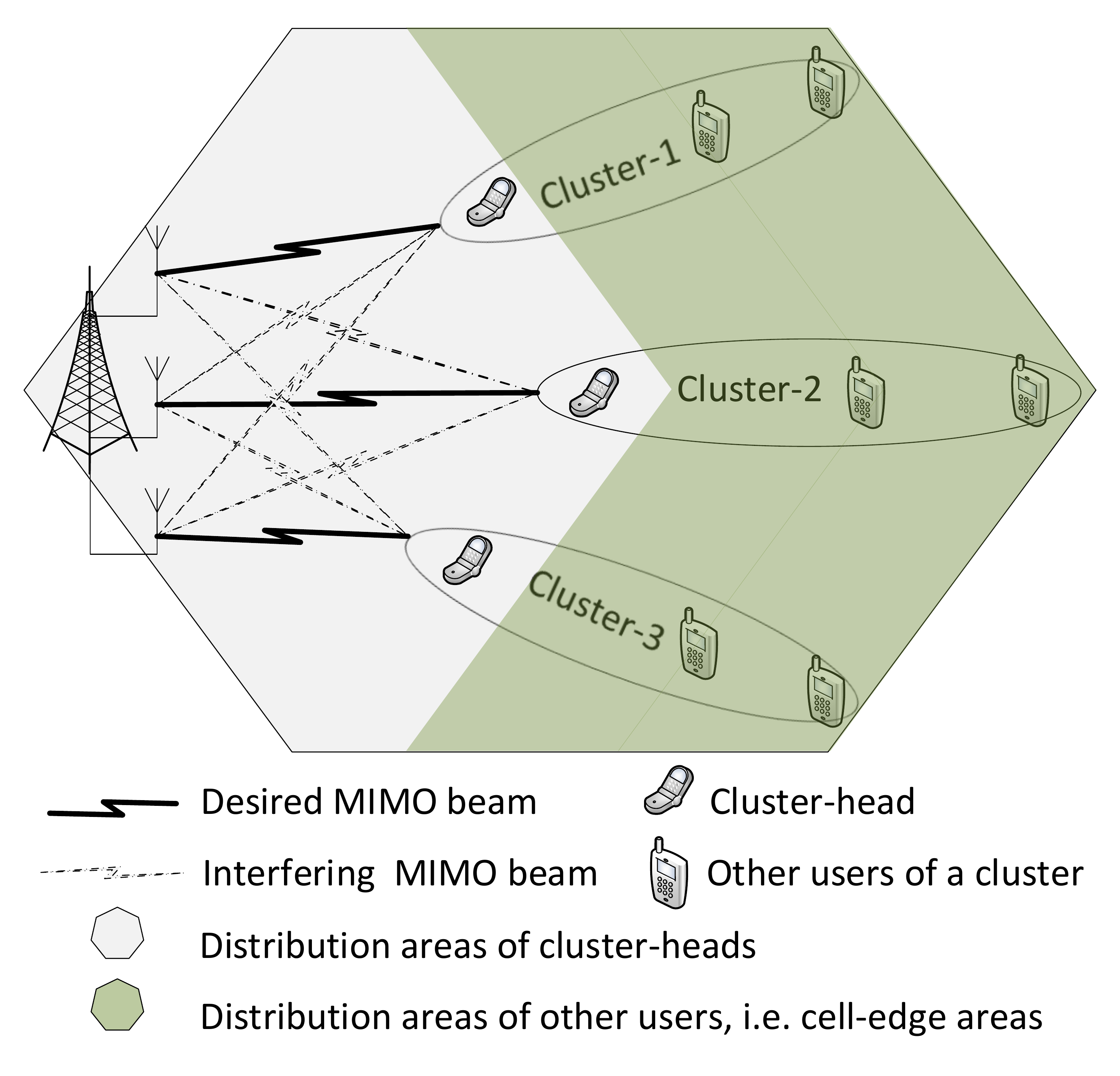}
	\caption{Illustrations of the simulation model for a $3$-user MIMO-NOMA system, where the number of the transmit antennas, receive antennas and UEs is $3, 9$, and $9$, respectively (i.e., $N_t = 3, L = 9, X =9$).}
	\label{fig:vi.a}
 \end{center}
\end{figure}

%======================================================================================================
\subsection{Simulation Results}
The performance evaluation of the proposed MIMO-NOMA system is performed at two different cases: uncorrelated Rayleigh fading channel gain users in each MIMO-NOMA cluster and correlated Rayleigh fading channel gain users in each MIMO-NOMA cluster.

%==========================================================================================================
\subsubsection{Users With Uncorrelated Channel Gains}
In a MIMO-NOMA system, all the users of a cluster might belong to different UEs. In addition, the throughput gain of the NOMA system increases if the channel gain differences among the participating users of a cluster increase. Therefore, it can be generally assumed  that the channel gains of the users in a MIMO-NOMA cluster are uncorrelated and identically and independently distributed (iid).

\begin{figure}[h]
\begin{center}
	\includegraphics[width=3.6 in]{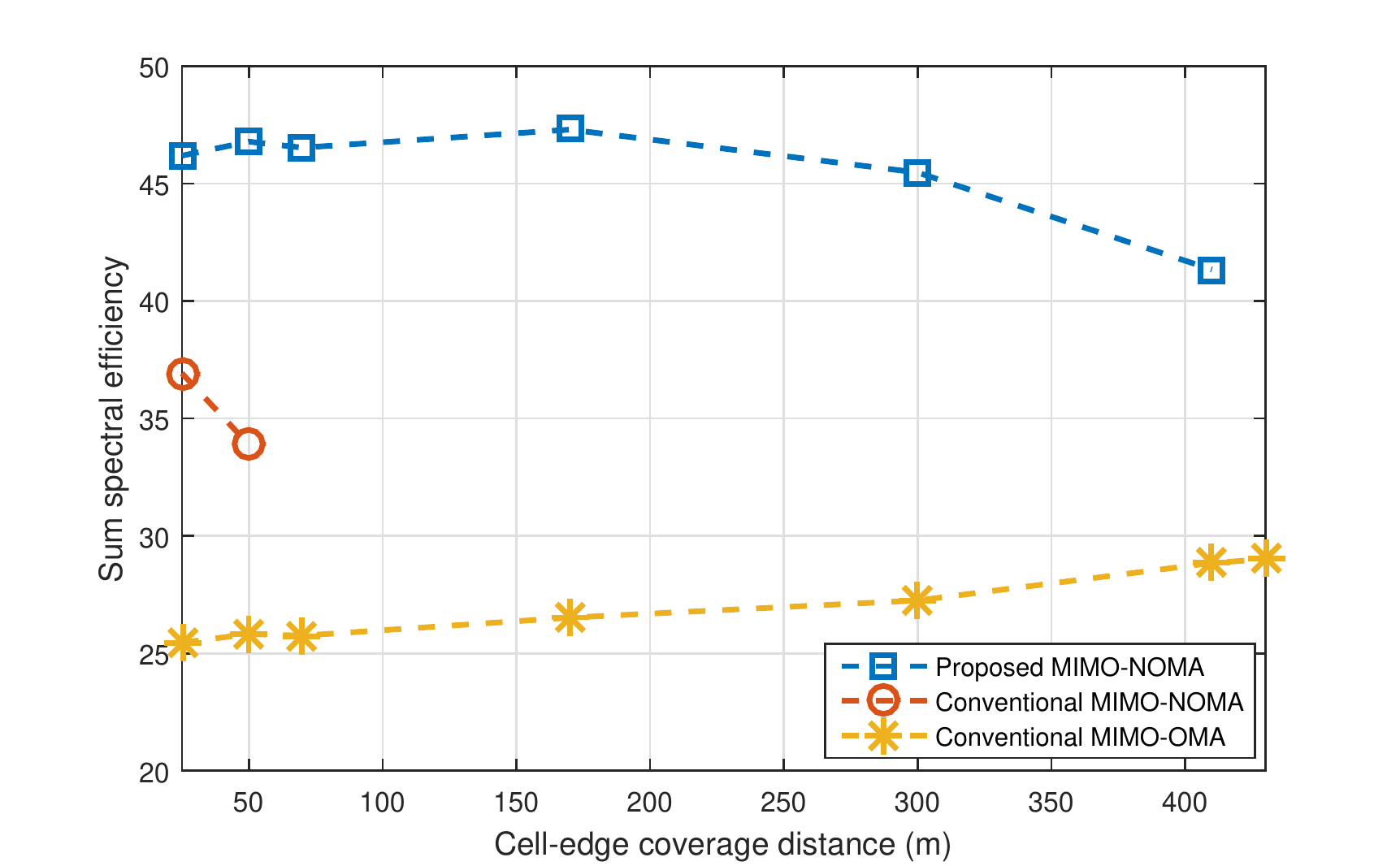}
	\caption{Spectral efficiency of a $2$-user MIMO-NOMA system for  $N_t = 3$ and $R_2 = $ OMA throughput with $50\%$ bandwidth. The cluster-heads are distributed within $150$m of the BS.}
	\label{fig:vi.1}
 \end{center}
\end{figure} 
\begin{figure}[h]
\begin{center}
	\includegraphics[width=3.6 in]{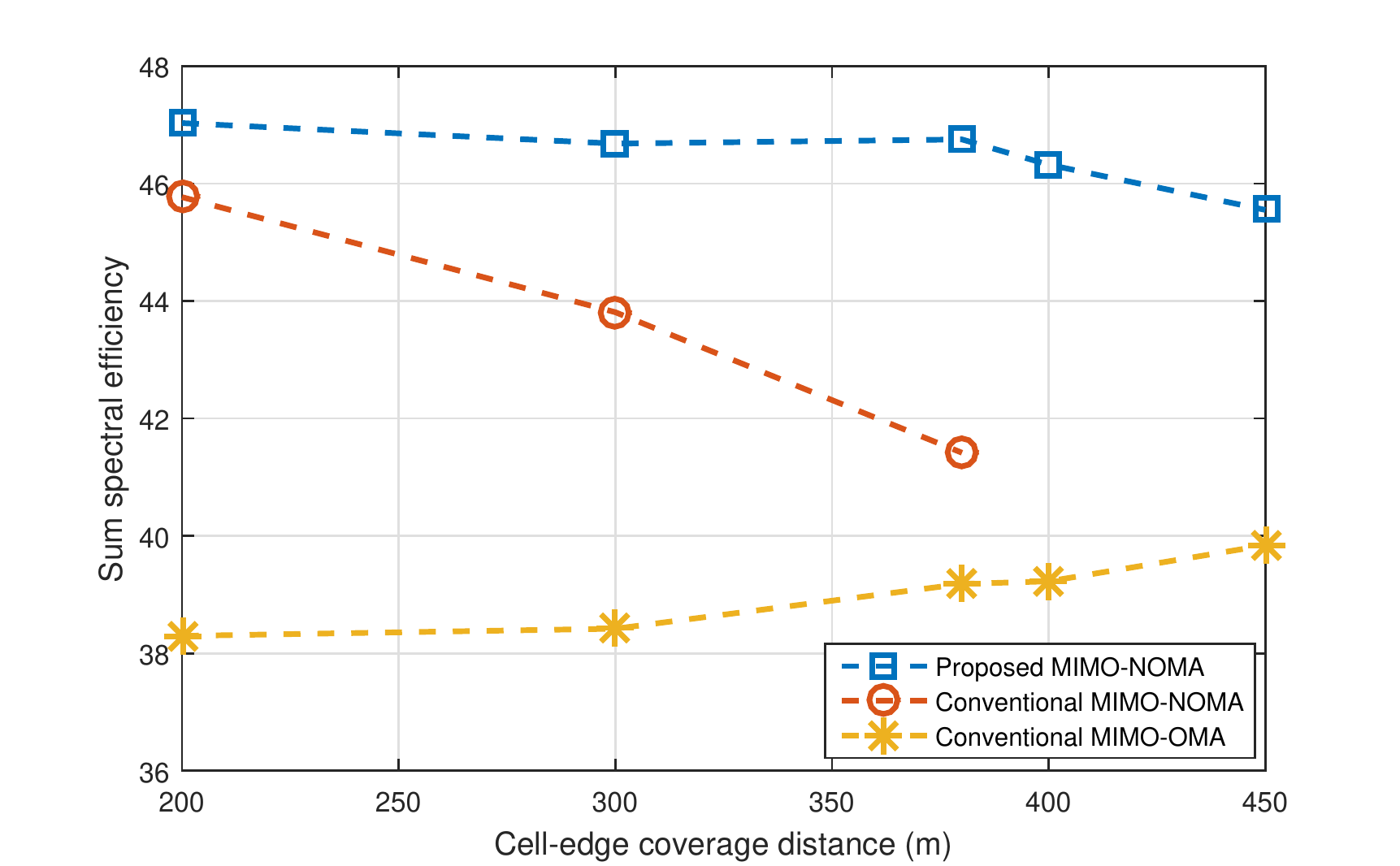}
	\caption{Spectral efficiency of a $2$-user MIMO-NOMA system for $N_t = 3$ and $R_2 = $ OMA throughput with $25\%$ bandwidth. The cluster-heads are distributed within $150$m of the BS.}
	\label{fig:vi.2}
 \end{center}
\end{figure} 
\begin{figure}[h]
\begin{center}
	\includegraphics[width=3.7 in]{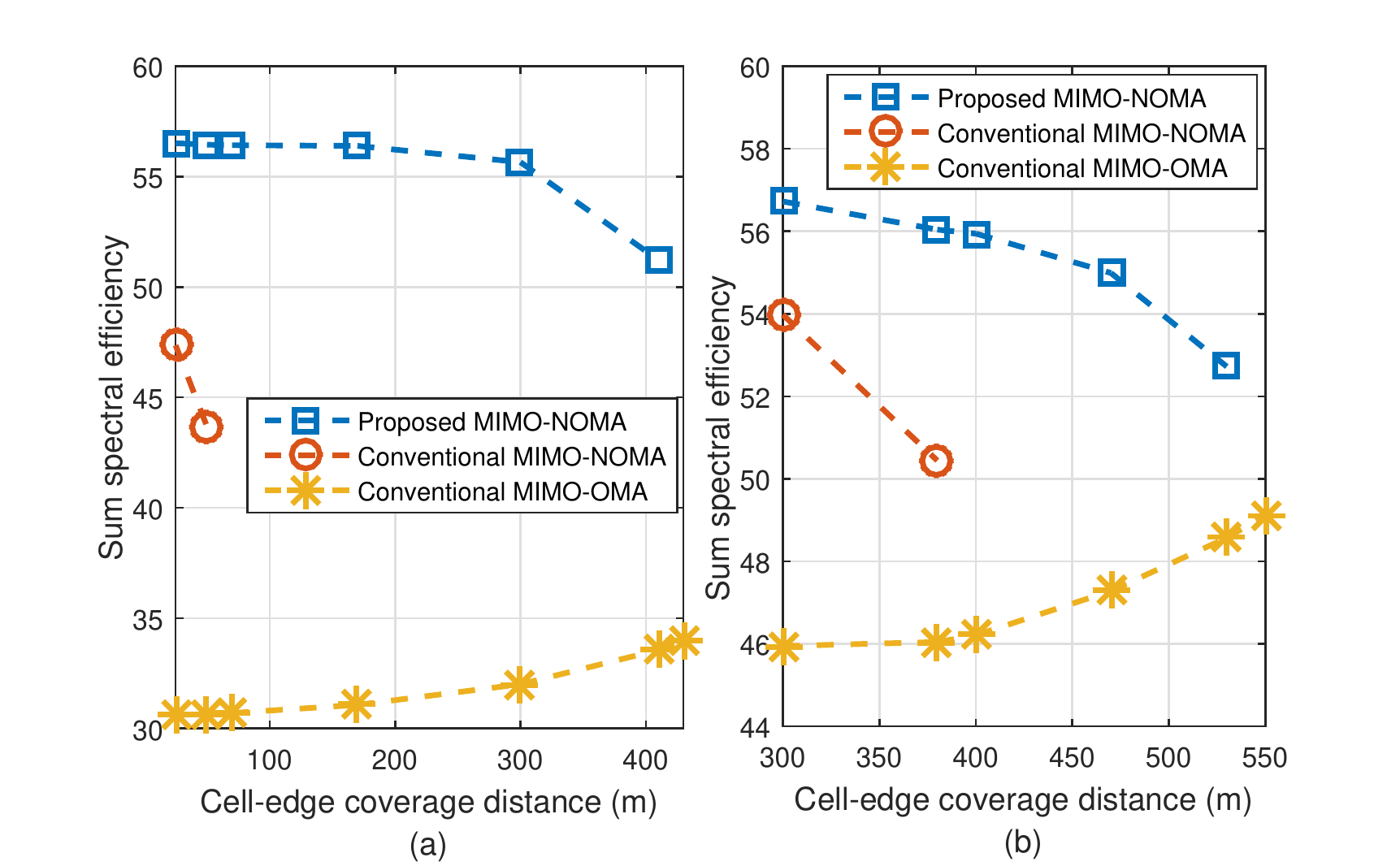}
	\caption{Spectral efficiency of a $2$-user MIMO-NOMA system for $N_t = 3$. (a) $R_2 = $ OMA throughput with $50\%$ bandwidth, (b) $R_2 = $ OMA throughput with $25\%$ bandwidth. The cluster-heads are distributed within $50$m of the BS.}
	\label{fig:vi.3}
 \end{center}
\end{figure} 

Figs. \ref{fig:vi.1}-\ref{fig:vi.2} show the sum spectral efficiencies for the proposed MIMO-NOMA, conventional MIMO-NOMA, and MIMO-OMA systems, where the guaranteed throughput requirement for the lower channel gain user of each $2$-user cluster is equal to its achievable OMA throughput by considering $50\%$ and $25\%$ of their cluster-bandwidth, respectively. The spectral efficiency performances in Figs. \ref{fig:vi.1}-\ref{fig:vi.2} are obtained by considering $3$ transmit antennas at BS end, while each MIMO-NOMA cluster contains $2$ users (i.e., $2$ single antenna UEs). The cluster-heads are randomly distributed within $150$m closer to the BS, while the lower channel gain users are distributed in the cell edge areas. The key observation from Figs. \ref{fig:vi.1}-\ref{fig:vi.2} is that the spectral efficiency (or throughput) gain of a MIMO-NOMA system is very high in comparison to that of a MIMO-OMA system, and the comparative gain of MIMO-NOMA over MIMO-OMA is much higher at higher guaranteed throughput requirements for lower channel gain users. These results are expected because high throughputs for the lower channel gain users require allocation of more spectrum resources  to them in an OMA system, while NOMA can simply meet such requirements by providing higher power to the lower channel gain users. However, at high throughput requirements, the lower channel gain users are more confined within limited cell-edge areas to form NOMA clusters (Fig. \ref{fig:vi.1}). In our proposed MIMO-NOMA model, as the lower channel gain users come closer to the cluster-heads, the inter-cluster interference increases. In such a case, the amount of transmit power required to meet their higher throughput requirements may not be available at the BS end.

Figs. \ref{fig:vi.3}(a)-\ref{fig:vi.3}(b) show spectral efficiency performances similar to those in Figs. \ref{fig:vi.1}-\ref{fig:vi.2}, respectively, for a scenario where the cluster-heads are randomly distributed within  $50$m closer to the BS. As can be observed from Fig. \ref{fig:vi.3},  the spectral efficiency of MIMO-NOMA is higher at higher throughput requirements of lower channel gain users, but  for such high data requirements, user clustering is limited by a certain coverage areas. In spite of similar performance trends, the  spectral efficiency gains shown in Fig. \ref{fig:vi.3} are much higher in comparison to those in  Figs. \ref{fig:vi.1}-\ref{fig:vi.2}. This performance enhancement is due to the improvement of the cluster-heads' channel gains since they are distributed  closer to the BS.  The results also show that the spectral efficiency of the proposed MIMO-NOMA system is much higher than that of a conventional MIMO-NOMA system thanks to the proposed precoding and decoding mechanisms which effectively reduces the inter-cluster interference.

\begin{figure}[h]
\begin{center}
	\includegraphics[width=3.7 in]{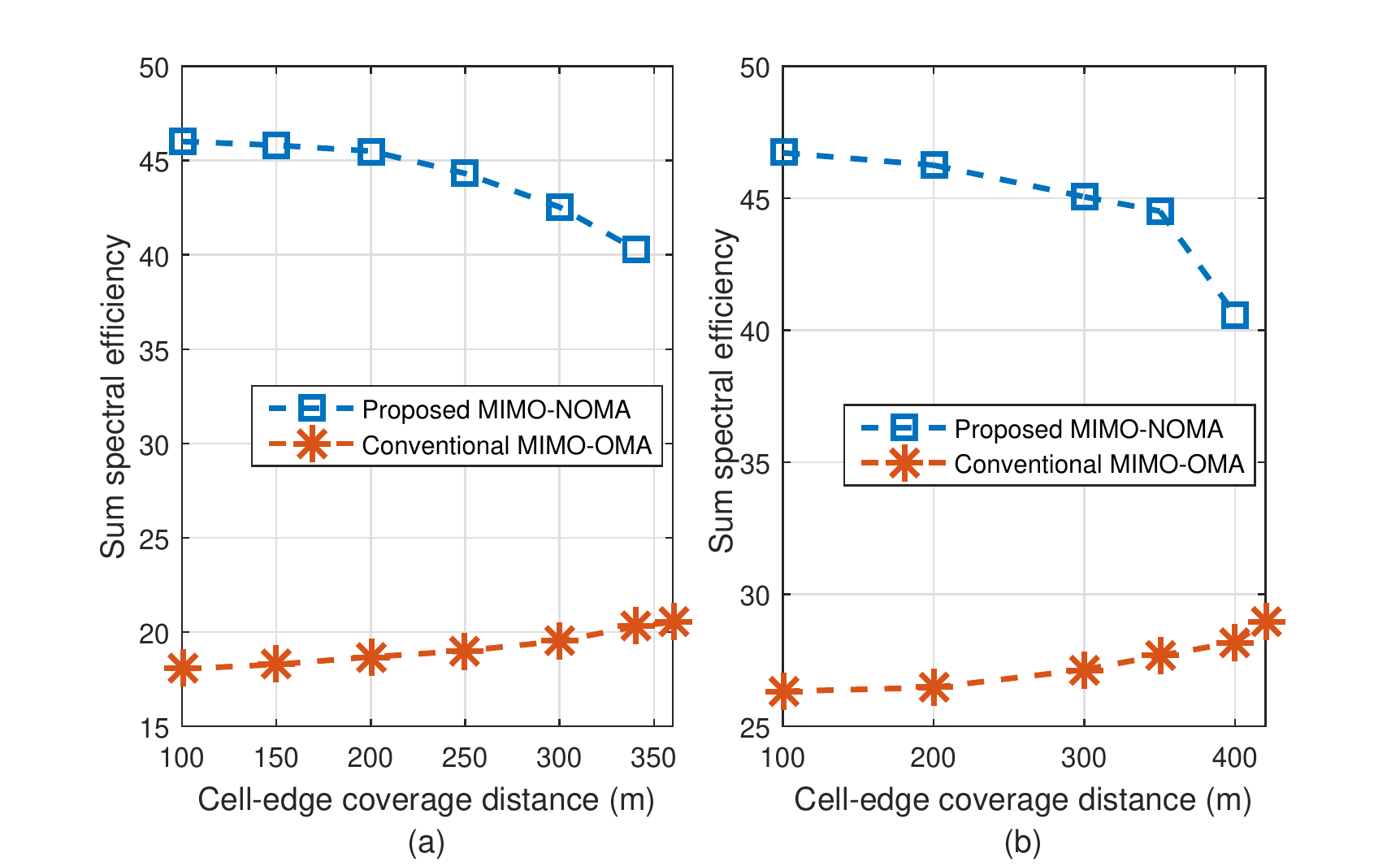}
	\caption{Spectral efficiency of a $3$-user MIMO-NOMA system for $N_t = 3$. (a) $R_i = $ OMA throughput with $33.33\%$ bandwidth, (b) $R_i = $ OMA throughput with $25\%$ bandwidth. The cluster-heads are distributed within $150$m of the BS ($i=2,3$).}
	\label{fig:vi.4}
 \end{center}
\end{figure} 
\begin{figure}[h]
\begin{center}
	\includegraphics[width=3.7 in]{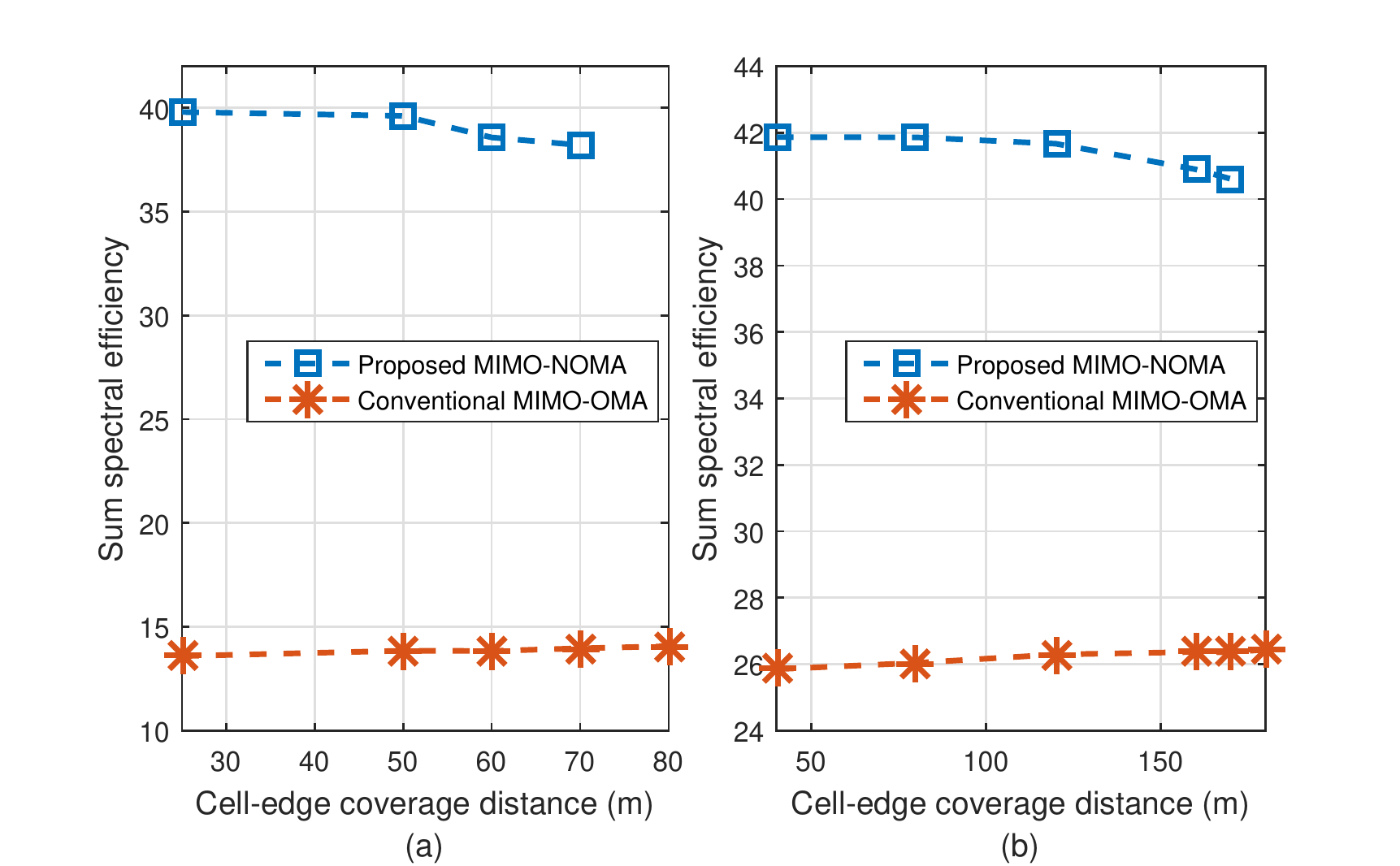}
	\caption{Spectral efficiency of a $4$-user MIMO-NOMA system for $N_t = 3$. (a) $R_i=$ OMA throughput with $25\%$ bandwidth, (b) $R_i =$ OMA throughput with $16.67\%$ bandwidth. The cluster-heads are distributed within $150$m of the BS ($i=2,3,4$).}
	\label{fig:vi.5}
 \end{center}
\end{figure} 
\begin{figure}[h]
\begin{center}
	\includegraphics[width=3.7 in]{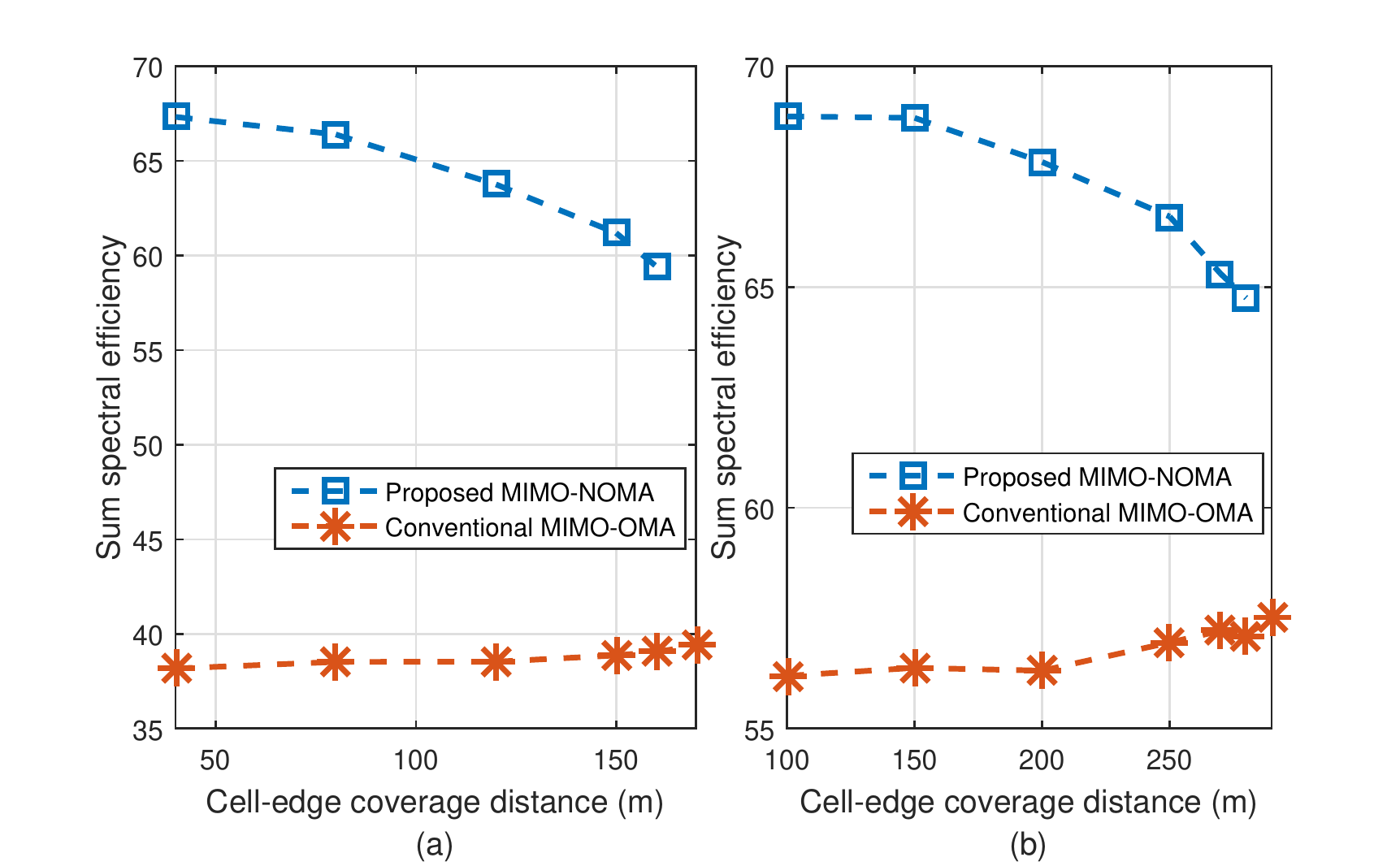}
	\caption{Spectral efficiency of a $2$-user MIMO-NOMA system for $N_t = 5$. (a) $R_2=$ OMA throughput with $50\%$ bandwidth, (b) $R_2 =$ OMA throughput with $25\%$ bandwidth. The cluster-heads are distributed within $150$m of the BS.}
	\label{fig:vi.6}
 \end{center}
\end{figure} 
\begin{figure}[h]
\begin{center}
	\includegraphics[width=3.7 in]{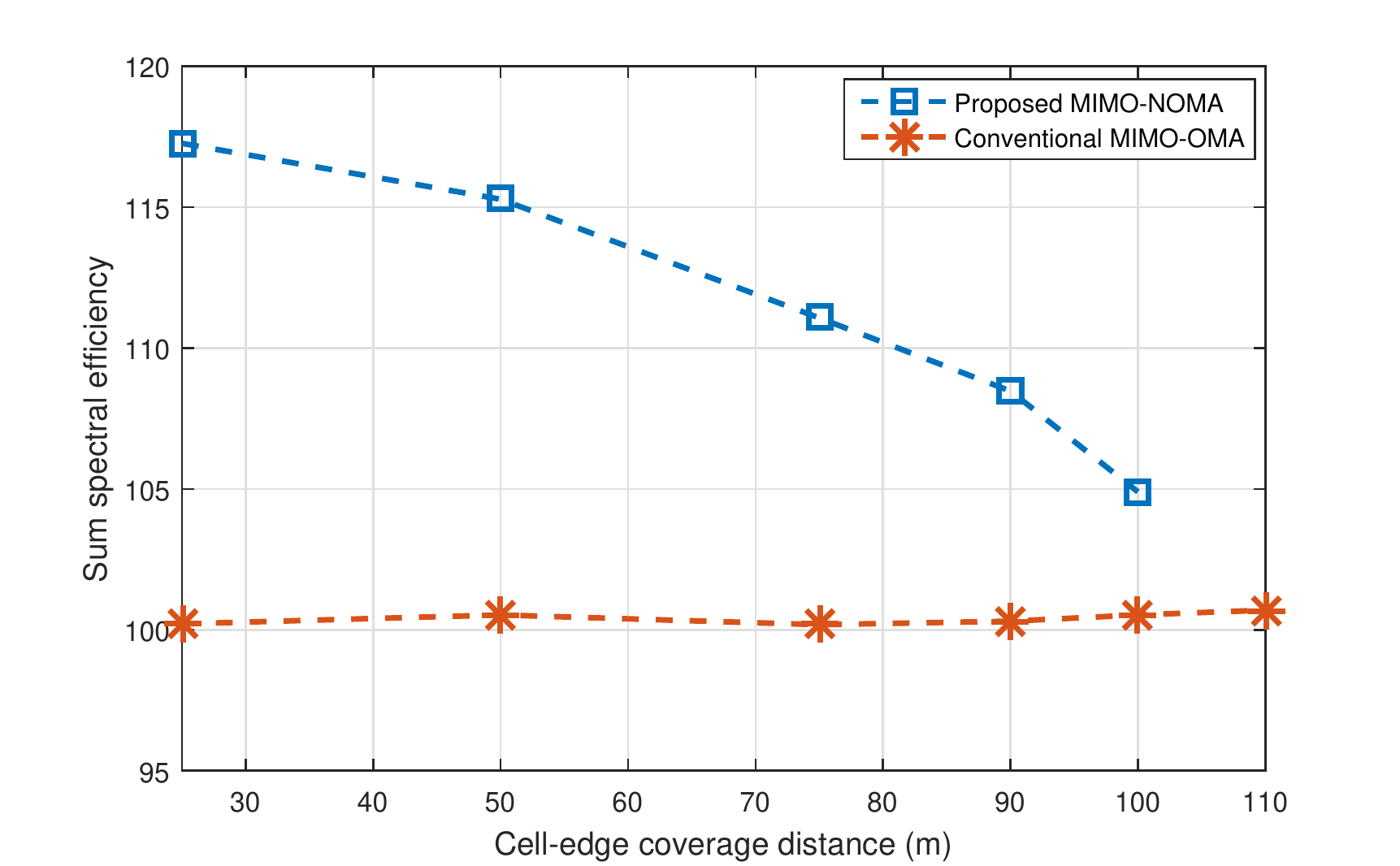}
	\caption{Spectral efficiency of a $3$-user MIMO-NOMA system for $N_t = 5$. $R_i =$ OMA throughput with $16.67\%$ bandwidth. The cluster-heads are distributed within $50$m of the BS ($i=2,3$).}
	\label{fig:vi.7}
 \end{center}
\end{figure} 
\begin{figure}[h]
\begin{center}
	\includegraphics[width=3.7 in]{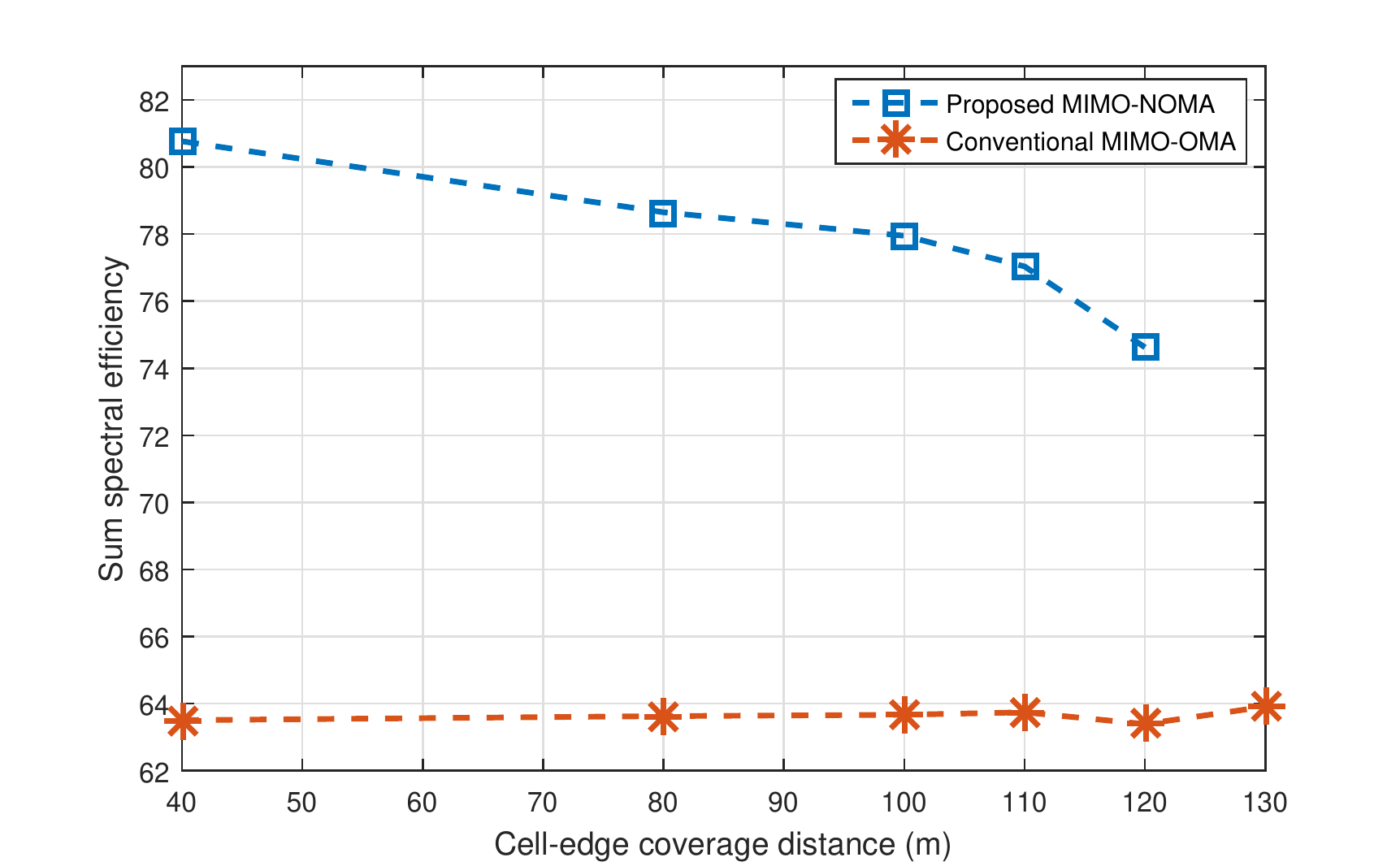}
	\caption{Spectral efficiency of a $2$-user MIMO-NOMA system for $N_t = 10$ and $R_2 =$ OMA throughput with $25\%$ bandwidth. The cluster-heads are distributed within $150$m of the BS.}
	\label{fig:vi.8}
 \end{center}
\end{figure} 
\begin{figure}[h]
\begin{center}
	\includegraphics[width=3.7 in]{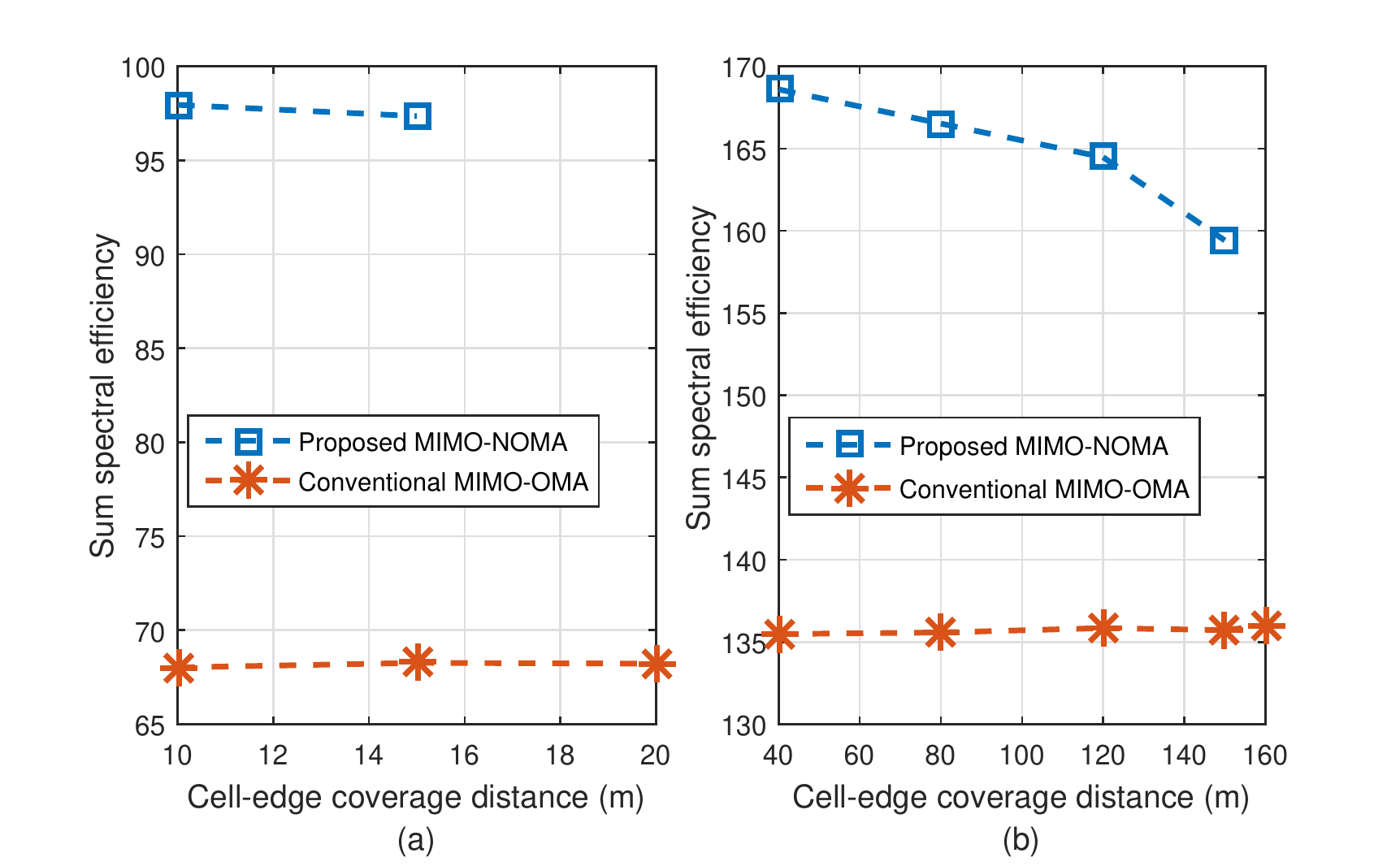}
	\caption{Spectral efficiency of a $2$-user MIMO-NOMA system for $N_t = 10$ with power control and $L=15$. The number of NOMA clusters is $5$, another $5$ users utilize OMA. (a) $R_2=$ OMA throughput with $50\%$ bandwidth, (b) $R_2 =$ OMA throughput with $25\%$ bandwidth. The cluster-heads are distributed within $150$m of the BS. For OMA spectral efficiency, all 15 users utilize OMA.}
	\label{fig:vi.9}
 \end{center}
\end{figure} 

For $3$-user and $4$-user clusters with $3$ transmit antennas at BS, the spectral efficiency performances of the proposed MIMO-NOMA and MIMO-OMA systems are presented in Fig. \ref{fig:vi.4} and Fig. \ref{fig:vi.5}, respectively. In Figs. \ref{fig:vi.4}-\ref{fig:vi.5}, the cluster-heads are assumed to be randomly distributed within $150$m closer to the BS and the other users are randomly distributed at the  cell edge areas. It is to be noted that, for these system parameters, the conventional MIMO-NOMA is unable to support more than $2$-user cluster. Fig. \ref{fig:vi.4}(a) and Fig. \ref{fig:vi.4}(b) are obtained, respectively, by considering the guaranteed throughput requirements for the lower channel gain users (i.e., users except the cluster-head) of each cluster to be equal to their achievable OMA throughput utilizing $33.33\%$ and $25\%$ of their cluster-bandwidth, respectively. Similar requirements for Fig. \ref{fig:vi.5}(a) and Fig. \ref{fig:vi.5}(b) are $25\%$ and $16.67\%$ of their cluster-bandwidth, respectively. By comparing all of the aforementioned results, it becomes evident that the comparative gain of the spectral efficiency of proposed MIMO-NOMA over the MIMO-OMA is much higher for higher order of user clusters, but the higher order user clusters would be confined within much smaller coverage areas. In addition, more users are required at the cell-edge areas to form higher order MIMO-NOMA systems, and the performance gain is improved if the data requirements for the cell-edge users are high. However, due to the strong inter-cluster interference and transmit power limitation, the proposed MIMO-NOMA cluster for more than $4$-user is not possible for the system parameters assumed for Fig. \ref{fig:vi.5}.

With $5$ transmit antennas at the BS, Figs. \ref{fig:vi.6}(a)-\ref{fig:vi.6}(b) show the spectral efficiencies of the $2$-user proposed MIMO-NOMA and the MIMO-OMA systems, where the throughput requirements and user distributions are similar to those used for Figs. \ref{fig:vi.1}-\ref{fig:vi.2}, respectively. Fig. \ref{fig:vi.6} shows the higher spectral efficiency for both of the MIMO-NOMA and MIMO-OMA systems in comparison of $2$-user cluster systems with $3$ transmit antennas at the BS (Figs. \ref{fig:vi.1}-\ref{fig:vi.3}). These gains are obtained due to the use of higher order MIMO. However, it is important to note that the inter-cluster interference of MIMO-NOMA increases in proportion to the number of transmit antennas at the BS end, i.e., for  higher order MIMO. Therefore, the coverage areas for the lower channel gain users are reduced for higher order MIMO scenarios, which can be clearly observed from Fig. \ref{fig:vi.6}. 

In high inter-cluster interference scenarios, higher order user clustering may not be possible to meet higher throughput requirements for lower channel gain users. Fig. \ref{fig:vi.7} shows the spectral efficiency of a $3$-user proposed MIMO-NOMA system and a MIMO-OMA system with $5$ transmit antennas at the BS. The throughput requirements for the lower channel gain users are equal to their achievable OMA throughputs by considering $16.67\%$ of their cluster-bandwidth. This $3$-user MIMO-NOMA is unable to meet the high throughput requirements by considering equal resource allocation among all the users in a cluster. With even more transmit antennas at the BS ($10$ transmit antennas), Fig. \ref{fig:vi.8} shows the spectral efficiency of the proposed MIMO-NOMA system and that of a MIMO-OMA system. 

%%In the following paragraph, what do you mean by power control for "single-user" clusters not using NOMA?

%% Answer: The users those are unable to use NOMA, they will be scheduled on OMA basis. I have mentioned these uers as single-user cluster. Since they are not scheduled on NOMA, thus transmit power control may not significantly affect their spectral efficiency if they experience good channel gains. While, In NOMA, the lower channel gain users need more power, and thus transmit power is the key resource for NOMA application. Any power control in NOMA system may significantly reduce the lower channel gain users spectral efficiency. Therefore, I suggest to perform power control to the OMA users but not NOMA users. 

The performance of NOMA with higher order MIMO is vulnerable to high inter-cluster interference. One straightforward way to reduce inter-cluster interference is to control the transmit powers of different non-orthogonal clusters. In higher order MIMO systems, only a portion of the randomly distributed users can use NOMA, while the others can be scheduled on an OMA basis. In such a scenario, some of the clusters would have only one user and transmit powers can be controlled for these single-user clusters. However,  the clusters utilizing NOMA should not be subjected to power control. With power control, Fig. \ref{fig:vi.9} shows the spectral efficiency of the proposed MIMO-NOMA and the MIMO-OMA systems with $10$ transmit antennas at BS. In fig. \ref{fig:vi.9}, five $2$-user clusters utilize $43$ dBm transmit power, while another five single-user clusters utilize $40$ dBm transmit power. Fig. \ref{fig:vi.9}(a) indicates the performance when the guaranteed throughput requirements of the NOMA users are equal to the their achievable OMA throughput for equally allocated cluster-bandwidth. Note that Fig. \ref{fig:vi.9}(b) is obtained by considering $25\%$ of the cluster-bandwidth for lower channel gain users' guaranteed throughput requirement.

%==============================================================================================
\subsubsection{Users with Correlated Channel Gains}
The correlated multi-path fading between the cluster-head and other users of a cluster is beneficial for MIMO-NOMA. In conventional MIMO-NOMA, the transmit beamforming is performed based on the channel information of cluster-head, and thus the other users of a MIMO-NOMA cluster can completely remove inter-cluster interference if their channel gains are fully correlated with that of the cluster-head. Our proposed MIMO-NOMA model also provides superior throughput performance under correlated channel gain scenarios. All the simulations under correlated channel gains are performed for $2$-user MIMO-NOMA clusters and the BS is assumed to be equipped with $5$ transmit antennas. 

\begin{figure}[h]
\begin{center}
	\includegraphics[width=3.7 in]{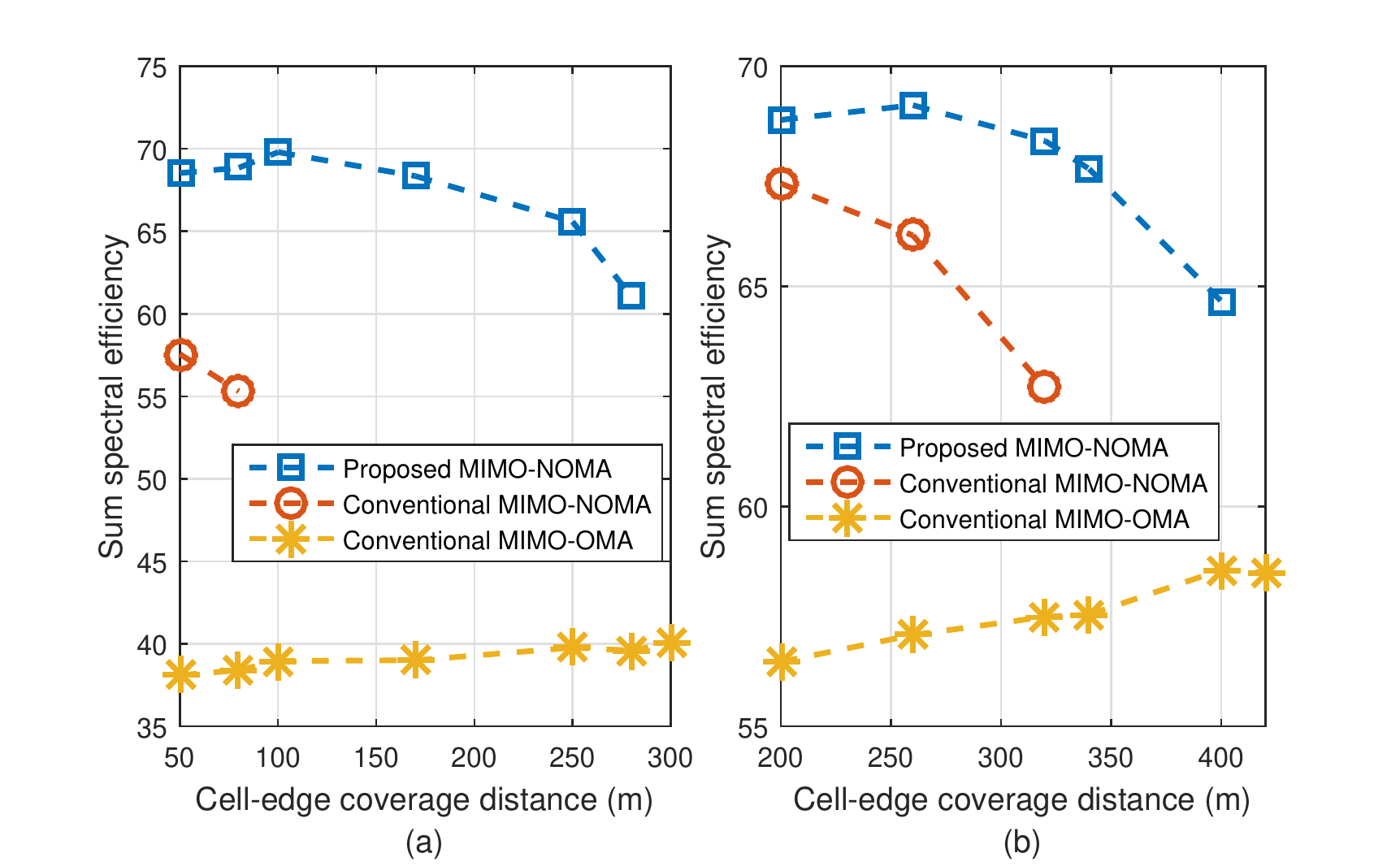}
	\caption{Spectral efficiency of a $2$-user MIMO-NOMA system for $\rho = 0.5$ and $N_t = 5$. (a) $R_2=$ OMA throughput with $50\%$ bandwidth, (b) $R_2 =$ OMA throughput with $25\%$ bandwidth. The cluster-heads are distributed within $150$m of the BS.}
	\label{fig:vi.10}
 \end{center}
\end{figure} 
\begin{figure}[h]
\begin{center}
	\includegraphics[width=3.7 in]{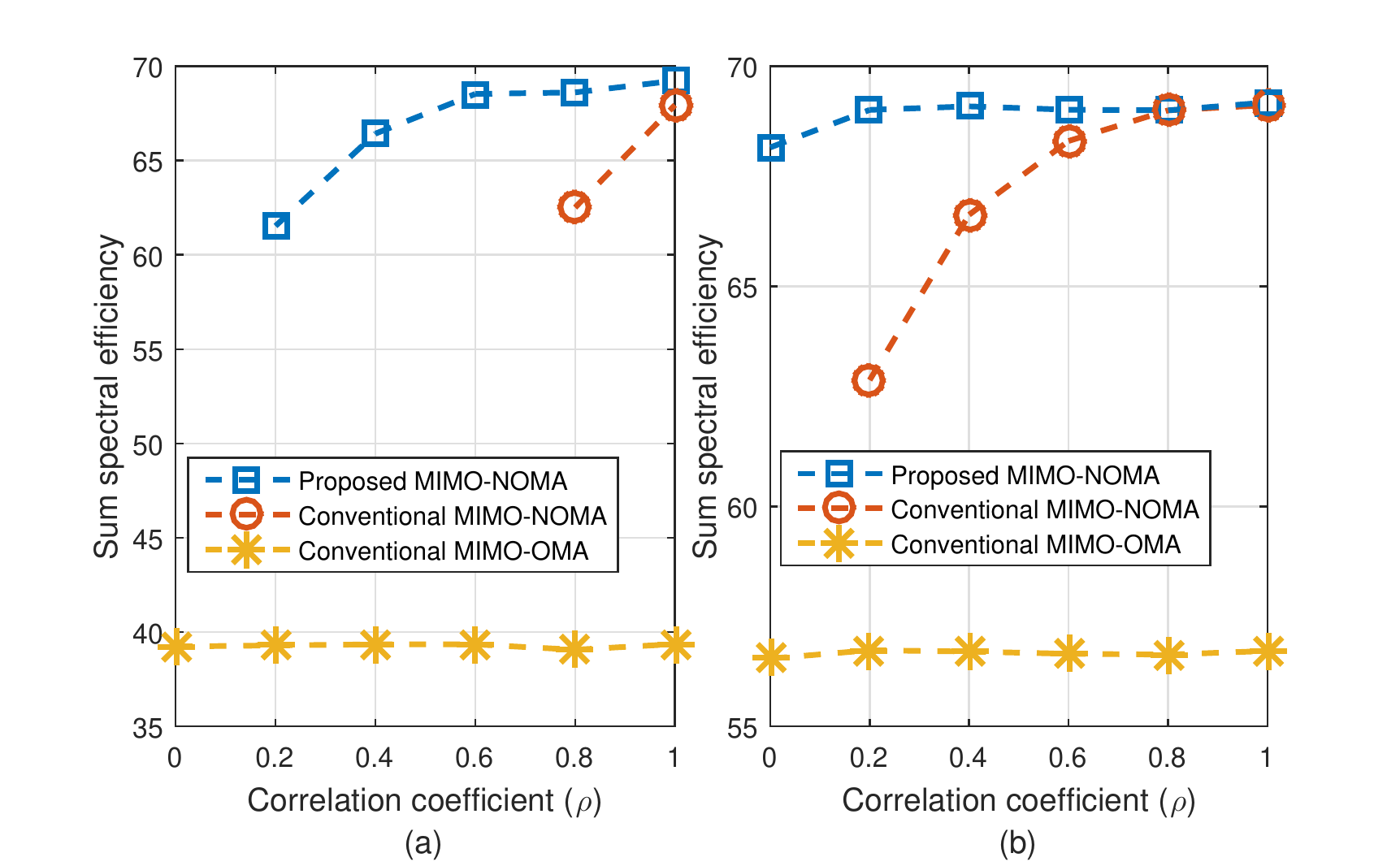}
	\caption{Spectral efficiency of a $2$-user MIMO-NOMA system for different values of $\rho$ for $N_t = 5$. (a) $R_2=$ OMA throughput with $50\%$ bandwidth, (b) $R_2 =$ OMA throughput with $25\%$ bandwidth. The cluster-heads are distributed within $150$m of the BS and other users are distributed within $200$m of cell edge.}
	\label{fig:vi.11}
 \end{center}
\end{figure} 

For a correlation coefficient of $\rho = 0.5$ between the Rayleigh fading channel gains of the cluster-head and that of each of the other users of a MIMO-NOMA cluster, Fig. \ref{fig:vi.10}(a) and Fig. \ref{fig:vi.10}(b) show the spectral efficiency of the MIMO-NOMA and MIMO-OMA for the guaranteed throughput requirements similar to Fig. \ref{fig:vi.1} and Fig. \ref{fig:vi.2}, respectively. Fig. \ref{fig:vi.10} clearly demonstrates the improvement of spectral efficiency of MIMO-NOMA (both for the proposed and conventional models) in comparison of their MIMO-OMA counterpart, while the proposed MIMO-NOMA model provides much higher spectral efficiency than the conventional MIMO-NOMA. Fig. \ref{fig:vi.10} is obtained by considering the cluster-heads' random distribution within $150$m closer to the BS, while other users are considered to randomly be distributed in the cell edge areas.

The spectral efficiencies for the MIMO-NOMA and MIMO-OMA systems for different correlation coefficient ($\rho = 0$ to $1$) are demonstrated in Fig. \ref{fig:vi.11}. The cluster-heads in Fig. \ref{fig:vi.11} are assumed to be randomly be distributed within $150$m closer to the BS, while the other users are considered to randomly be distributed within $200$m of cell edge distance. In addition, the guaranteed throughput requirements of the users considered in Fig. \ref{fig:vi.11}(a) and Fig. \ref{fig:vi.11}(b) are similar to those of the users considered in Fig. \ref{fig:vi.1} and Fig. \ref{fig:vi.2}, respectively. Fig. \ref{fig:vi.11} shows a huge improvement of the spectral efficiency of MIMO-NOMA system when highly correlated users are clustered. MIMO-NOMA clusters with very high correlated channel gains could completely cancel inter-cluster interference, thus the throughput performance is not affected by the number of transmit antennas. In such a channel condition, higher order MIMO-NOMA cluster (i.e., more users in a cluster) can provide higher throughput gains.

%====================================================================
\section{Conclusion}
%The combining of non-orthogonal multiple access (NOMA) and multiple input and multiple output (MIMO) technologies, known as MIMO-NOMA, is a promising approach for further enhancement of cell spectral efficiency. 

The application of MIMO-NOMA in wireless cellular system is a promising approach for enhancement of spectral efficiency performance. In this paper, the downlink multiuser MIMO-NOMA has been studied in which the total number of UE receive antennas in a cell is much higher than the number of BS transmit antennas. The receive antennas are grouped into a number of clusters and each cluster of MIMO-NOMA is served by a single MIMO beam which is orthogonal to the other clusters' beams, while all users in a cluster are scheduled on a NOMA basis. 

The use of MIMO-NOMA results in high inter-cluster interference, and the interference increases in proportion to the number of beams, i.e., the number of BS transmit antennas. In this paper, we have introduced a new inter-cluster zero-forcing beamforming technique for downlink MIMO-NOMA. The proposed technique can significantly eliminate inter-cluster interference when the more distinctive channel gain users are formed MIMO-NOMA clusters. We have also provided the dynamic power allocation solution for inter-cluster and intra-cluster power allocation in downlink MIMO-NOMA system. Moreover, a  user-clustering algorithm has been proposed which mitigates  inter-cluster interference  and helps achieving optimal intra-cluster power allocation in order to maximize the  overall throughput in a cell. Simulation results have demonstrated the spectral efficiency enhancement due to the proposed MIMO-NOMA model over the MIMO-OMA and other existing MIMO-NOMA solutions.  

The proposed MIMO-NOMA model has been evaluated in a single cell scenario.  A multi-cell system may introduce additional inter-cell interference which will affect the performance of the MIMO-NOMA system. A detailed investigation on the multi-cell MIMO-NOMA system is left as a future work.

%=======================================================================
%References
%=======================================================================
\bibliographystyle{IEEE}

\begin{thebibliography}{1}
%\bibitem{hossain2015}
%E. Hossain and M. Hasan, ``5G cellular: Key enabling technologies and research challenges,'' {\em IEEE Instrumentation and Measurement Magazine}, vol. 18, no. 3, June 2015, pp. 11--21.
%
%\bibitem{sesia2011}
%S. Sesia, I. Toufik, and M. Baker, {\em LTE$-$The UMTS Long Term Evolution: From Theory to Practice}. {\em 2nd Ed., Wiley}, 2011.

\bibitem{docomo2012}
NTT Docomo, ``Requirements, candidate solutions and technology roadmap for LTE Rel-12 onward,'' {\em 3GPP RWS-120010}, Jun. 2012.

\bibitem{saito2013}
Y. Saito, Y. Kishiyama, A. Benjebbour and T. Nakamura, ``Non-orthogonal multiple access (NOMA) for cellular future radio access,'' in {\em Proc. IEEE VTC Spring 2013}, Jun. 2013.

\bibitem{msali2016}
M.S. Ali, H. Tabassum and E. Hossain, ``Dynamic user clustering and power allocation in Non-orthogonal multiple access (NOMA) systems,'' in {\em IEEE Access J.}, early access 2016, pp. 01--17. (\textbf{Invited paper})

\bibitem{quentin2004}
Q.H. Spencer, A.L. Swindlehurst and M. Haardt, ``Zero-forcing methods for downlink spatial multiplexing in multiuser MIMO channels,''  in {\em IEEE Trans. Signal Process.}, vol. 52, no. 2, Feb. 2004, pp. 461--471.

\bibitem{kim2013}
B. Kim, S. Lim, H. Kim, S. Suh, J. Kwun, S. Choi, C Lee, s. Lee and D. Hong, ``Non-orthogonal multiple access in a downlink multiuser beamforming system,''  in {\em Proc. IEEE Military Commun. Conf.}, Nov. 2013, pp. 1278--1283.

\bibitem{higuchi2015}
K. Hoiguchi and A. Benjebbour, ``Non-orthogonal multiple access (NOMA) with successive interference cancellation for future radio access,'' {\em IEICE Trans. Commun.}, vol. E98--B, no.3, 2015, pp. 403--414. (\textbf{Invited paper})

\bibitem{higuch2013}
Y. Hayashi, Y. Kishiyama and K. Higuchi, ``Investigations on power allocation among beams in non-orthogonal access with random beam-forming and intra-beam SIC for cellular MIMO downlink,'' {\em Proc. IEEE Veh. Technol. Conf.}, Sep. 2013, pp. 1--5.

\bibitem{lan2014}
Y. Hayashi, Y. Kishiyama and K. Higuchi, ``Considerations on downlink non-orthogonal multiple access (NOMA) combined with closed-loop SU-MIMO,'' {\em Proc. IEEE Int. Conf. Signal Process. Commun. Syst.}, Dec. 2014, pp. 1--5.

\bibitem{ding2016}
Z. Ding, F. Adachi and H. V. Poor, ``The application of MIMO to non-orthogonal multiple access,'' {\em IEEE Trans. Wireless Commun.}, vol 15, no. 1, Jan. 2016, pp. 537--552.

\bibitem{qi2015}
Q. Sun, S. Han, Chin-Lin I, and Z. Pan, ``On the ergodic capacity of MIMO NOMA systems,'' {\em IEEE Wireless Commun. Lett.}, vol. 4, no. 4, Aug. 2015, pp. 405--408.

\bibitem{3GPP2006}
3GPP,``Physical layer aspects for evolved UTRA,'' {\em TR 25.814 (V7.0.0)}, Jun. 2006.

\end{thebibliography}

\end{document}